\documentclass[12pt]{article}
\usepackage{epsf}
\usepackage{cite}
\usepackage{amsmath,amssymb}
\input{colordvi.tex}
\usepackage[dvips]{graphicx}
\usepackage{comment}
\begin{document}

\newcommand{\lsim}{\stackrel{<}{_\sim}}
\newcommand{\gsim}{\stackrel{>}{_\sim}}

\newcommand{\rem}[1]{{$\spadesuit$\bf #1$\spadesuit$}}

\newcommand{\TT}[1]{\Red{#1}}

\renewcommand{\theequation}{\thesection.\arabic{equation}}
\renewcommand{\thefootnote}{\fnsymbol{footnote}}
\setcounter{footnote}{0}

\begin{titlepage}

\def\thefootnote{\fnsymbol{footnote}}

\begin{center}

\hfill UT-14-29\\
\hfill June, 2014\\

\vskip .75in

{\Large \bf 
  Studying Inflation with Future Space-Based
  \\ 
  Gravitational Wave Detectors
}

\vskip .75in

{\large Ryusuke Jinno$^{(a)}$, Takeo Moroi$^{(a)}$
  and Tomo Takahashi$^{(b)}$ }

\vskip 0.25in

{\em 
$^{(a)}$Department of Physics, The University of Tokyo,
Tokyo 113-0033, Japan
}

{\em
$^{(b)}$Department of Physics, Saga University, Saga 840-8502, Japan
}

\end{center}
\vskip .5in

\begin{abstract}

  Motivated by recent progress in our understanding of the $B$-mode
  polarization of cosmic microwave background (CMB), which provides
  important information about the inflationary gravitational waves
  (IGWs), we study the possibility to acquire information about the
  early universe using future space-based gravitational wave (GW)
  detectors.  We perform a detailed statistical analysis to estimate
  how well we can determine the reheating temperature after inflation
  as well as the amplitude, the tensor spectral index, and the running
  of the inflationary gravitational waves.  We discuss how the
  accuracies depend on noise parameters of the detector and the
  minimum frequency available in the analysis.  Implication of such a
  study on the test of inflation models is also discussed.

\end{abstract}

\end{titlepage}

\renewcommand{\thepage}{\arabic{page}}
\setcounter{page}{1}
\renewcommand{\thefootnote}{\#\arabic{footnote}}
\setcounter{footnote}{0}

\section{Introduction}
\setcounter{equation}{0}

The $B$-mode signal in the cosmic microwave background (CMB) provides
important information about the primordial inflation responsible for
the generation of the cosmic density fluctuations. In particular,
since the gravitational waves (GWs) produced during inflation is
likely to be the origin of the $B$-mode signal, the amplitude of the
inflationary gravitational waves (IGWs), at least for the scale
relevant for the CMB, will be understood once the $B$-mode signal is
observed.  Recently the discovery of the $B$-mode signal has been
announced by BICEP2 \cite{Ade:2014xna} and the reported
tensor-to-scalar ratio is relatively large (i.e., $r_{\rm
  BICEP2}=0.20^{+0.07}_{-0.05}$): such a value of the
  tensor-to-scalar ratio is consistent with the prediction of
  so-called large-field inflation like chaotic inflation
  \cite{Linde:1983gd}.  However, after the announcement of the BICEP2
  result, it was pointed out that the BICEP2 signal may be
  significantly affected by the polarized dust emission
  \cite{Mortonson:2014bja,Flauger:2014qra}.  Even in such a
  case, some fraction of the signal could originate from IGWs and
  hence a relatively large IGW amplitude may still be allowed.  The large
  value of the tensor-to-scalar ratio opens up a new possibility to
  detect and study the properties of the IGWs by direct detection
  experiments in the future \cite{Turner:1996ck, Pritchard:2004qp,
    Smith:2005mm, Cooray:2005xr, Smith:2006xf, Chongchitnan:2006pe,
    Friedman:2006zt, Smith:2008pf}.  Therefore, it would be
  interesting to consider to what extent we can obtain the information
  on the inflationary dynamics from future direct-detection
  experiments of GWs.\footnote
{ For future prospects of probing gravitational waves using $B$-mode
  in light of the BICEP2 result, see \cite{Caligiuri:2014sla}.  }



Importantly, the information about the dynamics of inflation is
imprinted in the IGWs.  In particular, the spectral index and the
running of the IGW spectrum depend on how the inflaton evolves during
inflation.  This fact implies that the determination of these
parameters from the IGWs may enable us to acquire the information
about the shape of the inflaton potential\cite{Seto:2005qy}.
Furthermore, the IGW spectrum is also sensitive to the history of the
universe so that the information about the cosmic expansion is
embedded in it.  In particular, the IGW spectrum changes its behavior
at the frequency corresponding to the time of the reheating due to the
inflaton decay \cite{Seto:2003kc}.\footnote
{For the works studying the thermal history with GWs along this line,
  see \cite{Nakayama:2008ip,Nakayama:2008wy,Kuroyanagi:2009br,
    Nakayama:2009ce,Kuroyanagi:2011fy,Kuroyanagi:2013ns}.}
If the spectrum of the IGWs is precisely studied, we may acquire
information about the very early epoch of the universe.

The possibilities of detecting and studying the IGWs with future
space-based GW detectors, like Big Bang Observer (BBO)\cite{Harry:2006fi} and 
DECi-hertz Interferometer Gravitational wave Observatory (DECIGO)
\cite{Seto:2001qf}, have been intensively studied.  In particular,
some of these detectors are expected to detect the IGW signal if
$r\sim 0.1$ which is predicted by the chaotic inflation (and also if
there is no significant suppression of the IGW amplitude at the
frequency range of $\sim 1\ {\rm Hz}$ compared to that at the CMB
scale).  In fact, the sensitivities of these detectors are planned to
be so high that they may not only detect the IGWs but also study their
properties.  Notably, these space-based GW detectors are sensitive to
the GWs at the frequency of $\sim 0.1-10\ {\rm Hz}$, which enters the
horizon when the cosmic temperature is about $10^7-10^9\ {\rm GeV}$.
Thus, we may have a chance to learn what happened in the universe at
such a high temperature with those GW detectors.  Since there will be
more data coming on the $B$-mode polarization of CMB in the near
future from Planck and other experiments, it would be now worth
revisiting the question of what kind of information we may acquire
with the future space-based GW detectors.

In this paper, we investigate how and how well we can study the
properties of the IGWs with future GW experiments.  For this purpose,
assuming future space-based GW experiments, we calculate the
signal-to-noise ratio for given fiducial models.  The size of the
signal is assumed to be the one predicted in the chaotic inflation,
while the noise functions are estimated for several choices of
detector parameters.  We first consider the measurements of IGW
parameters for the case where the reheating temperature is so high that the
IGW spectrum in the sensitivity range of the GW detectors is
insensitive to $T_{\rm R}$.  We study how well we can measure the
amplitude, the spectral index, and its running. Then, we discuss the
case where $T_{\rm R}$ is relatively low.

The organization of this paper is as follows.  In Section
\ref{sec:spectrum}, we briefly review basic properties of the IGW.
The statistical method we adopt in this paper is summarized in Section
\ref{sec:stat}.  The numerical results are shown in Section
\ref{sec:determination}.  In Section \ref{sec:test}, we discuss the
implication of the study of the IGW spectrum on the test of inflation
models.  In particular, we discuss how the prediction of the chaotic
inflation model can be tested.  Section \ref{sec:conclusions} is
devoted for conclusions and discussion.

\section{IGW Spectrum}
\label{sec:spectrum}
\setcounter{equation}{0}

We first briefly review basic properties of the IGWs for the case
where there is no entropy production after the decay of inflaton.  For
our analysis, it is convenient to define the present GW energy
density per log frequency normalized by the critical density $\rho_{\rm
  crit}$:
\begin{align}
  \Omega_{\rm IGW} (f) \equiv
  \frac{1}{\rho_{\rm crit}}
  \frac{d \rho_{\rm IGW}}{d \ln f},
\end{align}
where $f$ is the frequency of the GWs and $\rho_{\rm IGW}$ is the
total energy density of the IGW integrated over frequency.

The IGW spectrum $\Omega_{\rm IGW}$ strongly depends on the reheating
temperature after inflation.  In our analysis, we evaluate the
reheating temperature as\footnote
{Here, we assume the perturbative decay of the inflaton in the
    reheating process.  If the parametric resonance occurs, Eq.\
    \eqref{ReheatingTemp} is modified; in such a case, the reheating
    temperature is given by $T_{\rm R} \sim \sqrt{\Gamma_{\phi}^{\rm
        (P.R.)}M_{\rm Pl}}$, where $\Gamma_{\phi}^{\rm (P.R.)}$ is the
    particle-production rate due to the parametric resonance at the
    end of the reheating process.  Since the effect of the parametric
    resonance is model-dependent, hereafter, we consider the case
    where reheating process proceeds via the perturbative decay of the
    inflaton.}
\begin{align}
  T_{\rm R} \equiv
  \left(10 
    \frac{\Gamma_\phi^2 M_{\rm Pl}^2}{g(T_{\rm R}) \pi^2}
  \right)^{1/4},
  \label{ReheatingTemp}
\end{align}
with $\Gamma_\phi$ being the decay rate of the inflaton and $M_{\rm
  Pl}\simeq 2.4\times 10^{18}\ {\rm GeV}$ being the reduced Planck
mass. In addition, $g(T)$ is the effective number of massless degrees
of freedom at the temperature of $T$; we use the standard-model
prediction of $g(T\gg 100\ {\rm GeV})=106.75$, and $g(T_{\rm
  eq})=3.36$ (with $T_{\rm eq}$ being the temperature at the time of
radiation-matter equality).  We denote the frequency of the mode
entering the horizon at the time of the reheating as $f_{\rm R}$,
which is approximately given by
\begin{align}
  f_{\rm R} \simeq 0.3\ {\rm Hz} 
  \times \left( \frac{T_{\rm R}}{10^7\ {\rm GeV}} \right).
\end{align}

In the frequency range of our interest, $\Omega_{\rm IGW}$ is given by the
following form:
\begin{align}
  \Omega_{\rm IGW} (f) = 
  \bar{\Omega}_{\rm IGW} (f) {\cal T}(f),
  \label{OmegaGWtot}
\end{align}
where the function ${\cal T}(f)$ contains information about the reheating
temperature; ${\cal T}(f) \rightarrow 1$ as $f\ll f_{\rm R}$.  

We parameterize the primordial spectrum $\bar{\Omega}_{\rm IGW}$ by
introducing the amplitude, the tensor spectral index, and its running,
which are defined as
\begin{align}
  n_{\rm T} (f_*)  & \equiv
  \left[ \frac{d \ln \bar{\Omega}_{\rm IGW} (f)}{d \ln f} \right]_{f=f_*},
  \;\;\;
  \alpha_{\rm T} (f_*)  \equiv
  \left[ 
    \frac{d^2 \ln \bar{\Omega}_{\rm IGW} (f)}{d (\ln f)^2} 
  \right]_{f=f_*},
\end{align}
with $f_*$ being the pivot scale.  In slow-roll inflation models, the
parameters $n_{\rm T}$ and $\alpha_{\rm T}$ are expected to be much
smaller than $1$.  We will discuss how well we can constrain these
parameters with future space-based GW detectors.  It should be noted
that, because $\Omega_{\rm IGW} (f)$ does not depend on the pivot
scale, the following relations hold:
\begin{align}
  \ln \bar{\Omega}_{\rm IGW} (f'_*) & = 
  \ln \bar{\Omega}_{\rm IGW} (f_*) + n_{\rm T} (f_*) \ln (f'_*/f_*)
  + \frac{1}{2} \alpha_{\rm T} (f_*) \ln^2 (f'_*/f_*),
  \label{Omega'}
  \\
  n_{\rm T} (f'_*)  & = 
  n_{\rm T} (f_*) + \alpha_{\rm T} (f_*) \ln (f'_*/f_*),
  \label{nT'}
\end{align}
where we neglect the contributions of higher-order expansion
parameters.  Thus, the error in one parameter contaminates into those
of other parameters if we change the pivot scale.  In other words,
with a proper choice of $f_*$, the error of $\bar{\Omega}_{\rm IGW}
(f_*)$ (or $n_{\rm T} (f_*)$) can be minimized.  We will see that this
happens when $f_*$ is chosen to be the frequency at which the GW
detector has the best sensitivity.  We also note here that, if we
limit ourselves to second order in the expansion with respect to $\ln
(f/f_*)$, $\bar{\Omega}_{\rm IGW}$ is given in the following form:
\begin{align}
  \bar{\Omega}_{\rm IGW} (f) \simeq
  \bar{\Omega}_{\rm IGW} (f_*)  \left( 
    \frac{f}{f_*} 
  \right)^{n_{\rm T} (f_*) + \frac{1}{2} \alpha_{\rm T} (f_*) \ln (f/f_*)}.
  \label{OmegaGW}
\end{align}

As we have mentioned, the effect of the reheating is embedded in the
function ${\cal T}$.  We can understand the qualitative behavior of
${\cal T}$ by using the fact that the amplitude of the IGW is almost
constant when the wavelength is longer than the horizon scale while it
decreases as $a^{-2}$ (with $a$ being the scale factor) once it enters
the horizon.  For $f\gg f_{\rm R}$, ${\cal T}\propto f^{-2}$.  On the
contrary, for $f\ll f_{\rm R}$, ${\cal T}$ becomes close to $1$.  If
the reheating temperature is not high enough, a slight deviation from
the relation ${\cal T}=1$ may affect the determination of the IGW
parameters $n_{\rm T}$, and $\alpha_{\rm T}$.  When the universe is
dominated by the inflaton oscillation, the IGW amplitude behaves as
$\sim j_1(k\tau)/k\tau$, where $j_1$ is the spherical Bessel function,
$k$ is the conformal wavenumber, and $\tau$ is the conformal time.
Then, for $k\tau\ll 1$, which holds for superhorizon modes, the
evolution of the IGW amplitude has a slight dependence on $k$ as $\sim
1 + (k\tau)^2/10$, which results in a slight deviation from ${\cal T}=1$.  For
the mode with $f\ll f_{\rm R}$, we expect ${\cal T}(f)\simeq 1 +
c(f/f_{\rm R})^2$, with $c$ being a numerical constant.  We have
numerically calculated $c$, and found $c\simeq -0.3$.  Thus, if $f$ is
one or two orders of magnitude smaller than $f_{\rm R}$, the small
correction to ${\cal T}$ has minor effects on the measurements of
$n_{\rm T}$ and $\alpha_{\rm T}$ as far as they are of the order of
$10^{-2}-10^{-3}$.  For the IGW spectrum of $f\sim 1\ {\rm Hz}$, this
is the case when the reheating temperature is higher than $\sim
10^8-10^9\ {\rm GeV}$.  Because the effect of the reheating becomes
negligible when $f_{\rm R}$ is order of magnitude larger than the
frequency range relevant for the GW detectors, two types of analyses
are suggested.  One is the analysis with the assumption of high enough
reheating temperature; then we may impose ${\cal T}=1$ and determine
$\bar{\Omega}_{\rm IGW}$, $n_{\rm T}$, and $\alpha_{\rm T}$.  The
other is the one with $T_{\rm R}$ being included.  Then, we may have
information about the reheating temperature.  In Section
\ref{sec:determination}, we consider both cases.

\begin{figure}[t]
  \centerline{\epsfxsize=0.65\textwidth\epsfbox{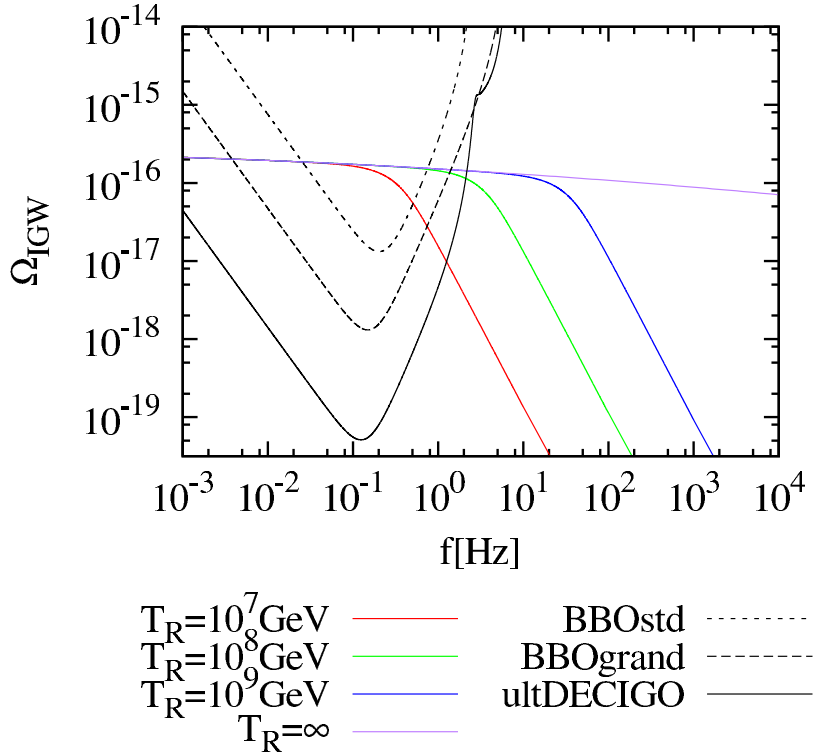}}
  \caption{\small Spectrum of the IGWs as a function of the frequency.
    Here, we take $\bar{\Omega}_{\rm IGW}=1.51\times 10^{-16}$,
    $n_{\rm T}=-6.38\times 10^{-2}$, and $\alpha_{\rm T}=-4.08\times
    10^{-3}$.  The reheating temperature is taken to be $T_{\rm
      R}=10^7\ {\rm GeV}$ (red), $T_{\rm R}=10^{8}\ {\rm GeV}$
    (green), $T_{\rm R}=10^{9}\ {\rm GeV}$ (blue), and high enough
    $T_{\rm R}$ (purple). The black lines are the effective
    sensitivity to the GW amplitude defined in Eq.\ \eqref{effGW}.}
  \label{fig:omegagw}
\end{figure}

In our analysis, ${\cal T}$ is evaluated by numerically solving the
evolution equation of GWs.  In Fig.\ \ref{fig:omegagw}, we show the
spectrum of the IGWs for several choices of parameters.  One can see a
significant suppression of $\Omega_{\rm IGW}$ in the high frequency
region.

With the inflation model being fixed, $\Omega_{\rm IGW}$ can be
evaluated.  In single-field slow-roll inflation model, 
$\bar{\Omega}_{\rm IGW}$ is given by
\begin{align}
  \bar{\Omega}_{\rm IGW} (f_*) = 
  \frac{1}{3} \Omega_{\rm rad} 
  \left( \frac{H_*}{2\pi M_{\rm Pl}} \right)^2
  \left( \frac{g(T_{f_*})}{g(T_{\rm eq})} \right)
  \left( \frac{g_{s}(T_{f_*})}{g_{s}(T_{\rm eq})} \right)^{-4/3},
  \label{BarOmg}
\end{align}
where $\Omega_{\rm rad}\simeq 9.4\times 10^{-5}$ is the density
parameter of radiation component, $H_*$ is the expansion rate of the
universe when the mode $f_*$ exits the horizon during inflation, and
$g_{s}(T)$ is the effective number of massless degrees of freedom for
entropy density at the temperature $T$.  (Here and hereafter, the
subscript ``$*$'' is used for quantities related to the mode with
$f=f_*$.)  In addition, $T_{f_*}\gg 100\ {\rm GeV}$ is the temperature
at the time of the horizon reentry of the mode $f_*$.  In the standard
model, $g_{s}(T\gg 100\ {\rm GeV})=106.75$, and $g_{s}(T_{\rm
  eq})=3.91$.  

The tensor spectral index $n_{\rm T}$ and its running $\alpha_{\rm T}$
are related to the so-called slow-roll parameters as
\begin{align}
  n_{\rm T} (f_*) & = - 2 \epsilon_*,
  \label{n_T}
  \\
  \alpha_{\rm T} (f_*) & = 
  - 2 \epsilon_* \left( 4 \epsilon_* - 2 \eta_* \right),
  \label{alpha_T}
\end{align}
where the quantities in the right-hand sides should be evaluated when
the mode with $f=f_*$ exits the horizon, and
\begin{align}
  \epsilon = \frac{1}{2} M_{\rm Pl}^2 \left( \frac{V'}{V} \right)^2,
  ~~~
  \eta = M_{\rm Pl}^2 \left( \frac{V''}{V} \right).
  \label{epseta}
\end{align}
Here $V$ is the potential of inflaton, with the ``prime'' being the
derivative with respect to the inflaton field.  We also note here that
the amplitude of the scalar-mode fluctuations is obtained as
\begin{align}
  A_{\rm S} =
  \frac{1}{2\epsilon}
  \left( \frac{H}{2\pi M_{\rm Pl}} \right)^2.
\end{align}

The purpose of the present study is to analyze the accuracy of the
determinations of $\bar{\Omega}_{\rm IGW}$, $n_{\rm T}$, $\alpha_{\rm
  T}$, and $T_{\rm R}$ in future space-based GW detectors.  The
accuracy, however, depend on the underlying (fiducial) values of these
parameters.  Here, we take the chaotic inflation model with a
quadratic potential \cite{Linde:1983gd} as an example and evaluate the
quantities introduced above.  (This model predicts the
tensor-to-scalar ratio of $r\simeq 0.15$, which is consistent with the BICEP2
observation.)  We adopt the inflaton potential of the
following form
\begin{align}
  V = \frac{1}{2} m_\phi^2 \phi^2.
  \label{infpot}
\end{align}
With the above inflaton potential, inflation occurs if the inflaton
$\phi$ starts its motion with the initial amplitude much larger than
the reduced Planck scale.  The evolution of the inflaton, as well as
that of the energy density of radiation, are governed by the following
equations:
\begin{align}
  & \ddot{\phi} + 3 H \dot{\phi} + m_\phi^2 \phi = 
  - \Gamma_\phi \dot{\phi},
  \label{eqphi}\\
  & \dot{\rho}_{\rm rad} + 4 H \rho_{\rm rad} =
  \Gamma_\phi \dot{\phi}^2,
  \label{eqrhor}
\end{align}
where the ``dot'' denotes the derivative with respect to time, 
$\rho_{\rm rad}$ is the energy density of radiation and $\Gamma_\phi$ is the decay rate of the inflaton.  We follow the
evolution of the universe by numerically solving the above equations.
Then, we calculate $\bar{\Omega}_{\rm IGW}$ by using Eq.\
\eqref{BarOmg}.  The value of $\bar{\Omega}_{\rm IGW}$, as well as
$A_{\rm S}$, depend on the inflaton mass $m_\phi$ and the reheating
temperature $T_{\rm R}$.  Here, we fix $m_\phi$ by requiring the amplitude for 
the scalar fluctuations 
$A_{\rm S}$ to satisfy \cite{Ade:2013zuv}:
\begin{align}
  A_{\rm S} (0.05\ {\rm Mpc}^{-1}) = 2.215 \times 10^{-9}.
\end{align}
For the reheating temperature of $T_{\rm R}=10^{7-12}\ {\rm GeV}$, the
best-fit value of $m_\phi$ is given by $(1.6-1.7)\times 10^{13}\ {\rm
  GeV}$.  In Table \ref{table:params}, we show the values of
$\bar{\Omega}_{\rm IGW}$, $n_{\rm T}$, and $\alpha_{\rm T}$ for
several values of the reheating temperature.  We also note here that,
if the inflaton interacts with the standard-model particles with
dimension-5 operator suppressed by the Planck scale, the decay rate of
the inflaton is roughly estimated to be
\begin{align}
  \Gamma_\phi \sim \frac{1}{4\pi} \frac{m_\phi^3}{M_{\rm Pl}^2}.
\end{align}
(Such a Planck suppressed interaction may arise if the cut-off
  scale of the standard model is around the Planck scale.)  Taking
$m_\phi\sim 10^{13}\ {\rm GeV}$, such a value of the decay rate
results in the reheating temperature of $\sim 10^{10}\ {\rm GeV}$.

\begin{table}[t]
  \begin{center}
    \begin{tabular}{l|ccc}
      \hline\hline
      $T_{\rm R}$ &
      $\bar{\Omega}_{\rm IGW}(1\ {\rm Hz})$ &
      $n_{\rm T}(1\ {\rm Hz})$ &
      $\alpha_{\rm T}(1\ {\rm Hz})$ \\
      \hline
      $10^7\ {\rm GeV}$ & 
      $1.41\times 10^{-16}$ &
      $-0.0751$ &
      $-0.00564$ \\
      $10^8\ {\rm GeV}$ & 
      $1.44\times 10^{-16}$ &
      $-0.0710$ &
      $-0.00504$ \\
      $10^9\ {\rm GeV}$ & 
      $1.48\times 10^{-16}$ &
      $-0.0672$ &
      $-0.00451$ \\
      $10^{10}\ {\rm GeV}$ & 
      $1.51\times 10^{-16}$ &
      $-0.0639$ &
      $-0.00408$ \\
      $10^{11}\ {\rm GeV}$ & 
      $1.54\times 10^{-16}$ &
      $-0.0609$ &
      $-0.00370$ \\
      $10^{12}\ {\rm GeV}$ & 
      $1.57\times 10^{-16}$ &
      $-0.0581$ &
      $-0.00337$ \\
      \hline\hline
    \end{tabular}
    \caption{The values of $\bar{\Omega}_{\rm IGW}$, $n_{\rm T}$, and
      $\alpha_{\rm T}$ at $1\ {\rm Hz}$ in the chaotic inflation model
      for several values of the reheating temperature.}
    \label{table:params}
  \end{center}
\end{table}

Before closing this section, we comment on our treatment of $n_{\rm
  T}$.  In the slow-roll single-field inflation model, the tensor
spectral index $n_{\rm T}$ is related to the tensor-to-scalar ratio
$r$ if these parameters are defined at the same wave-length (or
frequency). They are related as $n_{\rm T}=-\frac{1}{8}r$ at the
leading order of the slow-roll parameters.  However, this relation
hardly helps to fix $n_{\rm T}(f_*)$; experimental
determination of the tensor-to-scalar ratio at $f_*$ is difficult
since the information on the scalar-mode fluctuations at such a small
scale will not be available.  In addition, the value of $r$ at $f\sim f_*$
and that at the CMB scale (i.e., $\sim 0.05\ {\rm Mpc}^{-1}$) are
model-dependent and may significantly deviate.  For example, in the
chaotic inflation model, $r(0.05\ {\rm Mpc}^{-1})/r(k_*)\sim 3$.
Thus, we treat $n_{\rm T}(f_*)$ as one of the parameters which should
be determined.

\section{Statistical Analysis}
\label{sec:stat}
\setcounter{equation}{0}

Now we summarize how we estimate the underlying parameters which
govern the shape of the IGWs.  In the situation of our interest, the
data from the GW detector have information about the IGWs, while they
are also affected by the noise.  To reduce the effect of the noise,
the GW detectors with time-delay interferometry (TDI) method may be
used, which we assume in our analysis.  In particular, we concentrate
on two sets of spacecrafts at the vertices of (nearly) regular
triangles.  The first and second sets provide the TDI variables
so-called $(A,E,T)$ and $(A',E',T')$, respectively, which are
  linear combinations of the fluctuations of the laser
  frequency, normalized by the center one, measured at each spacecraft.  Written explicitly, each
  data stream $s_I(f)$ is
\begin{align}
s_A
&= \frac{1}{\sqrt{2}}(\alpha - \gamma), \\
s_E
&= \frac{1}{\sqrt{6}}(\alpha - 2 \beta + \gamma), \\
s_T
&= \frac{1}{\sqrt{3}}(\alpha + \beta + \gamma).
\end{align}
Here
\begin{align}
  \alpha
  &= y_{21}(t) - y_{31}(t) + y_{13}(t-L_2) - y_{12}(t-L_3) +
  y_{32}(t-L_1-L_2) - y_{23}(t-L_1-L_3),
\end{align}
with $y_{ij}$ being the normalized fluctuation of the laser frequency, 
propagating along arm $i$ (the one opposite to the spacecraft $i$) and
measured by spacecraft $j$ \cite{Tinto:2004wu, Estabrook:2000ef,
  Seto:2005qy}.  In addition, $\beta$ and $\gamma$ are obtained by the
cyclic permutations of the indices as $1\rightarrow 2\rightarrow
3\rightarrow 1$. The important point is that the noises of the
variables $A$, $E$, $T$ (and those of $A'$, $E'$, $T'$) are
uncorrelated.  Thus, with those variables, a high signal-to-noise
ratio may be realized.
Each data stream is given by the sum of the GW signal $H_I$
and the noise $n_I$:
\begin{align}
  s_I (f) = H_I (f) + n_I (f).
  \label{s_I}
\end{align}

The signal is linear in the amplitude of the
IGWs.  We expand the fluctuation of the metric for the tensor mode,
$h_{ij}=g_{ij}-\delta_{ij}$, as
\begin{align}
  h_{ij} = \sum_{P=+,\times} \int_{-\infty}^\infty df \int d \hat{n}
  h_P(f,\hat{n}) e^{2\pi i f (t-\hat{n}\vec{x})} \epsilon^P_{ij} (\hat{n}) ,
\end{align}
where $\epsilon^P_{ij}$ is the polarization tensor (whose
normalization is $\epsilon^P_{ij}\epsilon^{P'}_{ij}=2\delta_{PP'}$),
and $\hat{n}$ is the unit vector pointing to the direction of the
propagation. In the present convention,
\begin{align}
  \langle h_P^*(f,\hat{n}) h_{P'}(f,\hat{n}) \rangle =
  \frac{1}{8\pi} 
  \delta(f-f') \delta(\hat{n}-\hat{n}')
  \delta_{PP'} S_h(f,\hat{n}),
\end{align}
where $\langle\cdots\rangle$ denotes the ensemble average and
\begin{align}
  S_h(f) = \frac{3H_0^2}{4\pi^2} f^{-3} \Omega_{\rm IGW}(f),
\end{align}
with $H_0$ being the present Hubble parameter.  

\begin{figure}
  \begin{minipage}{0.5\columnwidth}
    \begin{center}
      \includegraphics[clip, width=1.0\columnwidth]{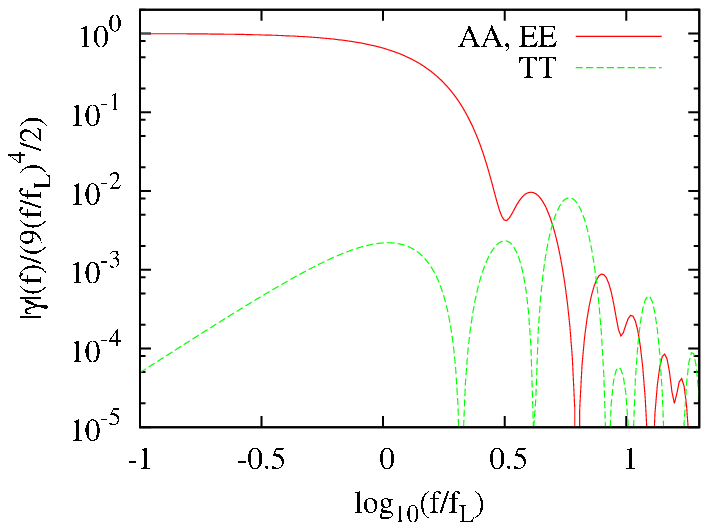}
    \end{center}
  \end{minipage}
  \begin{minipage}{0.5\columnwidth}
    \begin{center}
      \includegraphics[clip, width=1.0\columnwidth]{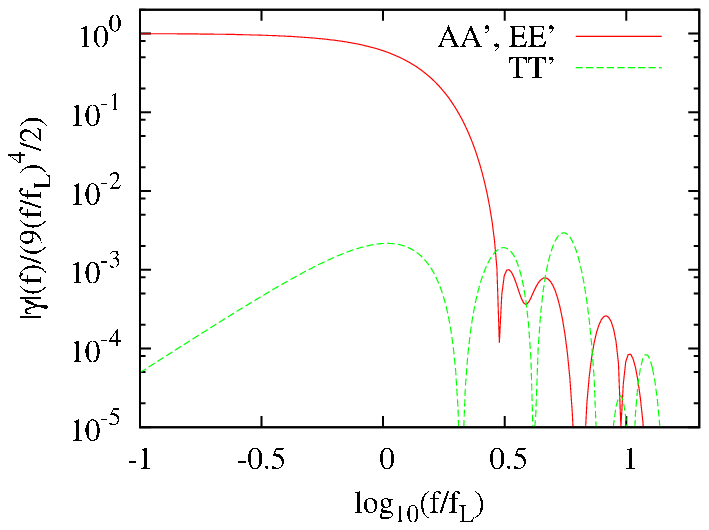}
    \end{center}
  \end{minipage}
  \caption{\small Overlap reduction function $\gamma_{II^{(')}}$
    for $I=A,E,T$. }
  \label{fig:overlap}
\end{figure}

Because the IGW amplitude is so small that the signal can be well
  approximated to be proportional to $h_P$, the GW signal can be
  expressed in the following form:
\begin{align}
  H_I (f) = \sum_{P=+,\times} \int d \hat{n}
  R_I (f,\hat{n}, P) h_P(f,\hat{n}).
\end{align}
Here, the information about the detector geometry is
embedded into the function $R_I$ \cite{Corbin:2005ny},
with which the overlap reduction function can be obtained
(see Eq.\ \eqref{eq:overlap} below).
Then, the two-point correlator of the signal becomes
\begin{align}
  \langle H_I^* (f) H_{J} (f') \rangle =
  \frac{\gamma_{IJ}}{5}
  \delta (f-f') S_h(f),
\end{align}
where the overlap reduction function is given by
\begin{align}
  \gamma_{IJ} \equiv
  \frac{5}{2} \sum_{P=+,\times} \int \frac{d \hat{n}}{4\pi}
  R_I^* (f,\hat{n}, P) R_{J} (f,\hat{n}, P).
  \label{eq:overlap}
\end{align}
In Fig.\ \ref{fig:overlap}, we plot $\gamma_{IJ}$ for several choices
of $(I,J)$.\footnote{ For the calculation of the overlap reduction
  function, see \cite{Cornish:2001bb,Corbin:2005ny}.}

For the calculation of the data stream given in Eq.\ \eqref{s_I}, the
noise power spectrum is defined as
\begin{align}
  \langle n_I^*(f) n_{I}(f') \rangle
  = \frac{1}{2} \delta(f-f') S_I(f).
\end{align}
The functional forms of the spectra are given by
\cite{Prince:2002hp,Nayak:2002ir}
\begin{align}
  S_A(f)
  = & 8\sin^2(f/2f_L) \nonumber \\
   &\times \left[
    \left(2+\cos(f/f_L) \right) S_y^{\rm optical-path} + 2\left(3+2\cos(f/f_L) + \cos(2f/f_L) \right) S_y^{\rm proof-mass}
  \right], 
  \label{S_A} \\
  S_E(f)
  =& S_A(f), \\
  S_T(f) 
  =& 2\left( 1+2\cos(f/f_L)\right)^2
  \left[S_y^{\rm optical-path}+4\sin^2(f/2f_L)S_y^{\rm proof-mass} \right],
  \label{S_T}
\end{align}
where $f_L=c/2\pi L$ with $L$ being the arm length, and $S_y^{\rm optical-path}$
and $S_y^{\rm proof-mass}$ parameterize the effects of the optical-path noise and
proof-mass noise, respectively.  Here, we adopt the noise spectrum
for the BBO-standard (abbreviated as BBO-std) and BBO-grand given in
\cite{Seto:2005qy}.  
In order to see the result for the ideal situation, we also
consider the case where the sensitivity is only limited by the
standard quantum limit.  In such a case, we adopt the arm length of
$5\times 10^7\ {\rm m}$ and the mass of $100\ {\rm kg}$, and determine 
$S_y^{\rm optical-path}$ dominates $S_y^{\rm proof-mass}$ at the frequency range of
$f\gtrsim 0.1\ {\rm Hz}$, assuming that only shot noise contributes to the former; 
our reference values are $S_y^{\rm optical-path}=4.7 \times
10^{-51} (f/{\rm Hz})^2 \; {\rm Hz}^{-1}$ and $S_y^{\rm proof-mass}=4.7 \times
10^{-55} (f/{\rm Hz})^{-2} \; {\rm Hz}^{-1}$.  (We call this case as
ultimate-DECIGO, abbreviated as ult-DECIGO.)  The values of $S_y^{\rm optical-path}$ and $S_y^{\rm proof-mass}$ 
used in our analysis are summarized in Table\ \ref{table:noise}.

\begin{table}[t]
  \begin{center}
    \begin{tabular}{l|ccc}
     \hline\hline
     {Experiments} & $L[{\rm m}]$ & $S_y^{\rm optical-path}[(f/{\rm Hz})^2 \; {\rm Hz}^{-1}]$ 
     &$S_y^{\rm proof-mass}[(f/{\rm Hz})^{-2} \; {\rm Hz}^{-1}]$\\
      \hline
      {BBO-std} & 
      $5\times 10^7$ &
      $3.6\times 10^{-49}$ &
      $2.5 \times 10^{-52}$ \\
      {BBO-grand} & 
      $2\times 10^7$ &
      $1.2\times 10^{-50}$ &
      $2.5 \times 10^{-54}$ \\
      {ult-DECIGO} & 
      $5\times 10^7$ &
      $4.7 \times 10^{-51}$ &
      $4.7 \times 10^{-55}$\\
      \hline\hline
    \end{tabular}
    \caption{Parameters for the noise power spectra.}
    \label{table:noise}
  \end{center}
\end{table}

In our analysis, we use the information about the cross correlation
for $(I,I')=(A,A')$, $(E,E')$, and $(T,T')$.  To see the sensitivities
of each experimental setup, we first calculate the spectrum of the
IGWs for a given set of fiducial parameters $\{ \hat{p} \}$. (In this paper, 
the ``hat'' is used for fiducial values.) 
Using the fact that the noises are uncorrelated for these sets of cross
correlations, we calculate $\delta\chi^2$ by postulating the
values of the fundamental parameters $\{
p\}$\cite{Kudoh:2005as}:
\begin{align}
\label{eq:delta_chi2}
  \delta \chi^2  (\{ p \}; \{ \hat{p} \}) = 
  - 2 \ln {\cal L} (\{ p \}; \{ \hat{p} \})
   = 
   \frac{2}{25}
   T_{\rm obs} \sum_{(I,I')}
   \int_{f_{\rm min}}^{\infty} df
   \frac{\gamma_{II'}^2 (f) }
   {{\sigma}_{II'}^2 (f)}
   \left[
     S_h(f; \{ p \}) 
     - S_h(f; \{ \hat{p} \})
   \right]^2.
\end{align}
Here, $S_h(f; \{ \hat{p} \})$ and $S_h(f; \{ p \})$ are calculated
with the fiducial parameters $\{ \hat{p} \}$ and the postulated
parameters $\{ p \}$, respectively, $T_{\rm obs}$ is the observation
time, and
\begin{align}
  {\sigma}_{II'}^2 (f) = 
  \left[ \frac{1}{2}S_I(f)+\frac{1}{5}\gamma_{II}(f)S_h(f) \right]
  \left[ \frac{1}{2}S_{I'}(f) + \frac{1}{5}\gamma_{I'I'}(f)S_h(f) \right] 
  + \frac{1}{25}\gamma_{II'}^2(f) S_h^2(f).
  \label{sigmaII'}
\end{align}
We note here that the stochastic cosmic background GWs for $f\lesssim
O(0.1\ {\rm Hz})$ are expected to be dominated by GWs from white-dwarf
binaries, and hence the low-frequency data may not be used for the
study of the IGWs.  We introduce the minimum frequency $f_{\rm min}$
to take this fact into account.  We will discuss how the result
changes as we vary $f_{\rm min}$.  We can also calculate the
signal-to-noise ratio $S/N$ for detection as
\begin{align}
  (S/N)^2 
   & = 
   \frac{2}{25}
   T_{\rm obs} \sum_{(I,I')}
   \int_{f_{\rm min}}^{\infty} df
   \frac{\gamma_{II'}^2 (f) }
   { {{\sigma}_{II'}^{\rm (null)}}^2 (f)}
   \left[
      S_h(f; \{ \hat{p} \})
   \right]^2,
\end{align}
where ${\sigma}_{II'}^{\rm (null)}$ is calculated with the assumption
of the null signal,
\begin{align}
  {{\sigma}_{II'}^{\rm (null)}}^2 (f) = 
  \frac{1}{4} S_I(f) S_{I'}(f).
\end{align}
Also, we define the effective strain sensitivity \cite{Cornish:2001bb}
\begin{align}
h_{\rm eff}^{-2}(f)
&= \left( \frac{2}{25}T_{\rm obs}f \sum_{(I,I')} \frac{\gamma_{II'}^2(f)}{{{\sigma}_{II'}^{\rm (null)}}^2 (f)} \right)^{1/2},
\label{heff}
\end{align}
and the effective sensitivity to the GW amplitude $\Delta \Omega_{\rm IGW}$
\begin{align}
\Delta \Omega_{\rm IGW}(f)
&= \frac{4\pi^2}{3H_0^2} f^3 h_{\rm eff}^2(f),
\label{effGW}
\end{align}
with which
\begin{align}
  (S/N)^2 &= 
  \int_{f_{\rm min}}^{\infty} d\ln f 
  \frac{\Omega_{\rm IGW}^2(f;\hat{p})}{\Delta \Omega_{\rm IGW}^2(f)}.
\end{align}
In Fig.\ \ref{fig:omegagw}, we also plot $\Delta \Omega_{\rm IGW}$ with $T_{\rm obs}=1{\rm yr}$ for
each experiment.

For some fundamental parameters, the likelihood function can be approximated by 
Gaussian. We collectively denote such parameters by $\{ p_a \}$, while 
denote the other parameters by $\{ p_A \}$. 
(Here and hereafter, in the subscripts, small letters are used for Gaussian parameters, while 
capital letters are for non-Gaussian ones.)
Then Eq.~\eqref{eq:delta_chi2} can be expanded as
\begin{align}
\delta \chi^2  (\{ p_a, p_A \}; \{ \hat{p} \}) 
&= \delta \chi^2  (\{ p_a^{\rm (ref)}, p_A \}; \{ \hat{p} \}) 
+ \sum_{ab}
 (p_a - p_a^{\rm (ref)}) (p_b - p_b^{\rm (ref)})
 {\mathcal F}_{ab} (\{ p_a^{\rm (ref)}, p_A \}),
 \label{deltachi2_expand}
\end{align}
where ${\mathcal F}_{ab} (\{ p_a^{\rm (ref)}, p_A \})$ is the Fisher matrix
\begin{align}
 {\mathcal F}_{ab} (\{ p_a^{\rm (ref)}, p_A \})
   & = 
  \frac{2}{25}
  T_{\rm obs} \sum_{(I,I')}
  \int_{f_{\rm min}}^{\infty} df
  \frac{\gamma_{II'}^2(f)\partial_{p_a}S_h(f;\{ p_a^{\rm (ref)}, p_A \})\partial_{p_b}S_h(f;\{ p_a^{\rm (ref)}, p_A \})}
  {{\sigma}_{II'}^2 (f)},
 \label{Fisher}
\end{align}
with $\partial_{p_a}$ being the derivative with respect to the fundamental parameter $p_a$, 
and $\{ p_a^{\rm (ref)} \}$ is the value of $\{ p_a \}$ which gives the minimum of $\delta \chi^2$ 
for given $\{ p_A \}$.
In the following analysis we take 
$\ln \bar{\Omega}_{\rm IGW}$, $n_{\rm T}$, $\alpha_{\rm T}$ and $T_{\rm R}$ 
as fundamental parameters which describe IGWs. 
As we will see, the likelihood function for the reheating temperature $T_{\rm R}$ cannot be
approximated by Gaussian in some cases, thus we take 
$\{ p_a \} = \{ \ln \bar{\Omega}_{\rm IGW}, n_{\rm T}, \alpha_{\rm T} \}$ and 
$\{ p_A \} = \{ T_{\rm R} \}$. 

To study the expected constraints on the fundamental parameters, 
we evaluate the likelihood function by
using Eq.~\eqref{deltachi2_expand}.  In presenting constraints on above
mentioned parameters in two-dimensional plane or one-dimensional axis,
we marginalize over irrelevant parameters, which we 
denote by $p_\perp$ collectively, by integrating $p_\perp$ as
\begin{align}
 \tilde{\cal L} (\{ p \}; \{ \hat{p} \}) =
 \int dp_\perp {\cal L} (\{ p \}; \{ \hat{p} \}).
 \label{L-tilde}
\end{align}
(It should be understood that, in the first argument $\{ p \}$ of
$\tilde{\cal L} (\{ p \}; \{ \hat{p} \})$, $p_\perp$ is not included.)
Then, the $\delta\chi^2$ can be obtained as
\begin{align}
 \delta \chi^2 (\{ p \}; \{ \hat{p} \}) \equiv
 -2 \ln 
 \frac{\tilde{\cal L} (\{ p \}; \{ \hat{p} \}) }
 {\tilde{\cal L} (\{ \hat{p} \}; \{ \hat{p} \}) }.
\end{align}

\section{Determination of the IGW Spectrum}
\label{sec:determination}
\setcounter{equation}{0}

Now we consider how well we can probe the properties of IGWs
using future space-based GW detectors.  Throughout this paper, we
consider the simplest scenario in which there is no extra entropy
production after the decay of inflaton.\footnote
{If there exists extra entropy production after the reheating, the IGW
  spectrum may show significant change compared to the standard case.
  For example, if there occurred a phase transition in the early
  universe, the IGW spectrum may show characteristic feature at the
  frequency corresponding to the time of the cosmic phase transition
  \cite{Jinno:2011sw, Jinno:2013xqa}. The case with late-time entropy
  production has also been studied in \cite{Nakayama:2009ce,
    Kuroyanagi:2013ns}. In any case, we do not consider such a
  possibility in this paper.}

In the following subsections, we show the results of our analysis for
the cases with and without including the reheating temperature into
the list of fit parameters.  In our numerical calculation, we choose
the following as fundamental parameters which determine the IGW
spectrum:
\begin{align*}
  \ln \bar{\Omega}_{\rm IGW}(f_*),~~~
  n_{\rm T}(f_*), ~~~
  \alpha_{\rm T}(f_*),~~~
  T_{\rm R}.
\end{align*}
(Hereafter we sometimes omit the argument $f_*$ for notational simplicity.) 
In the following, $f_*$ is optimized in each analysis. 
(Note that the expected
uncertainties in the determination of the fundamental parameters do
not change even if we vary the fiducial values within the range shown
in Table \ref{table:params}.)  In calculating likelihood 
${\mathcal L}$, we assume it to be Gaussian in $\bar{\Omega}_{\rm
  IGW}$, $n_{\rm T}$ and $\alpha_{\rm T}$ directions, while we do not
in $T_{\rm R}$ direction and use Eq.~\eqref{deltachi2_expand}.

We use the predictions of the chaotic inflation model with $T_{\rm
  R}=10^{10}\ {\rm GeV}$ as the fiducial values, irrespective of the
fiducial value of the reheating temperature to make the comparison
easier.  Then, at $1\ {\rm Hz}$, 
\begin{align}
  \hat{\bar{\Omega}}_{\rm IGW} (1\ {\rm Hz}) = 1.51\times 10^{-16},~~~
  \hat{n}_{\rm T} (1\ {\rm Hz}) =-0.064, ~~~
  \hat{\alpha}_{\rm T}  (1\ {\rm Hz})  =-0.0041.
  \label{at1Hz}
\end{align}
The amplitude, the tensor spectral index, and its running at $f=f_*$ are
evaluated based on \eqref{at1Hz}. 

\subsection{Case with high enough $T_{\rm R}$}

First, we study the accuracy of the measurements of the parameters
$\bar{\Omega}_{\rm IGW}$, $n_{\rm T}$, and $\alpha_{\rm T}$, assuming
the shape of the IGW spectrum given in Eq.\ \eqref{OmegaGW} and ${\cal
  T}(f)\rightarrow 1$.  This is relevant if the reheating temperature
is so high that $f_{\rm R}$ is much larger than the frequency relevant
for the GW detectors.  We note here that the large tensor-to-scalar
ratio recently reported by BICEP2, together with the scalar amplitude
and the $e$-folding, is consistent with the chaotic inflation model
with the mass scale $m_\phi$ of order $10^{13}\ {\rm GeV}$.  Such a
value of $m_\phi$ results in the reheating temperature as high as
$\sim 10^{10}\ {\rm GeV}$ if the inflaton couples to the
standard-model sector via Planck-suppressed operator, which gives
$f_{\rm R}$ much higher than $1\ {\rm Hz}$.  Thus, for such a case,
the analysis given in this subsection is relevant.

\begin{figure}
  \begin{center}

  \begin{minipage}{0.4\columnwidth}
    \begin{center}
      \includegraphics[clip, width=1.0\columnwidth]{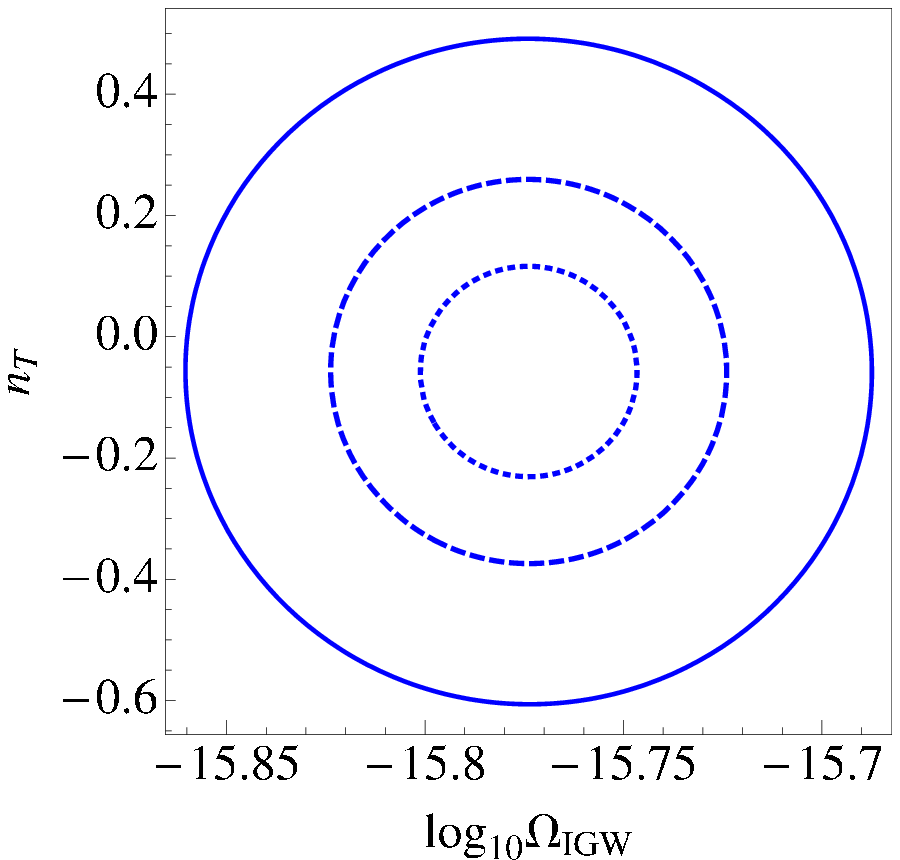}
    \end{center}
  \end{minipage}
  \begin{minipage}{0.4\columnwidth}
    \begin{center}
      \includegraphics[clip, width=1.0\columnwidth]{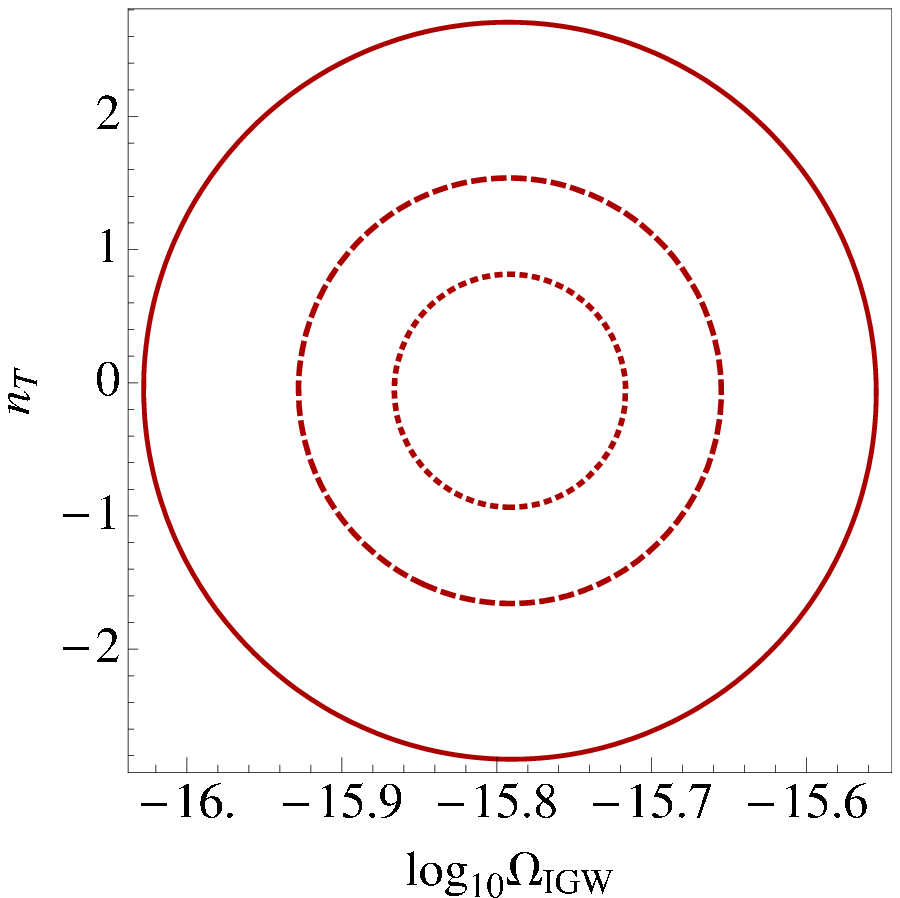}
    \end{center}
  \end{minipage}

  \begin{minipage}{0.4\columnwidth}
    \begin{center}
      \includegraphics[clip, width=1.0\columnwidth]{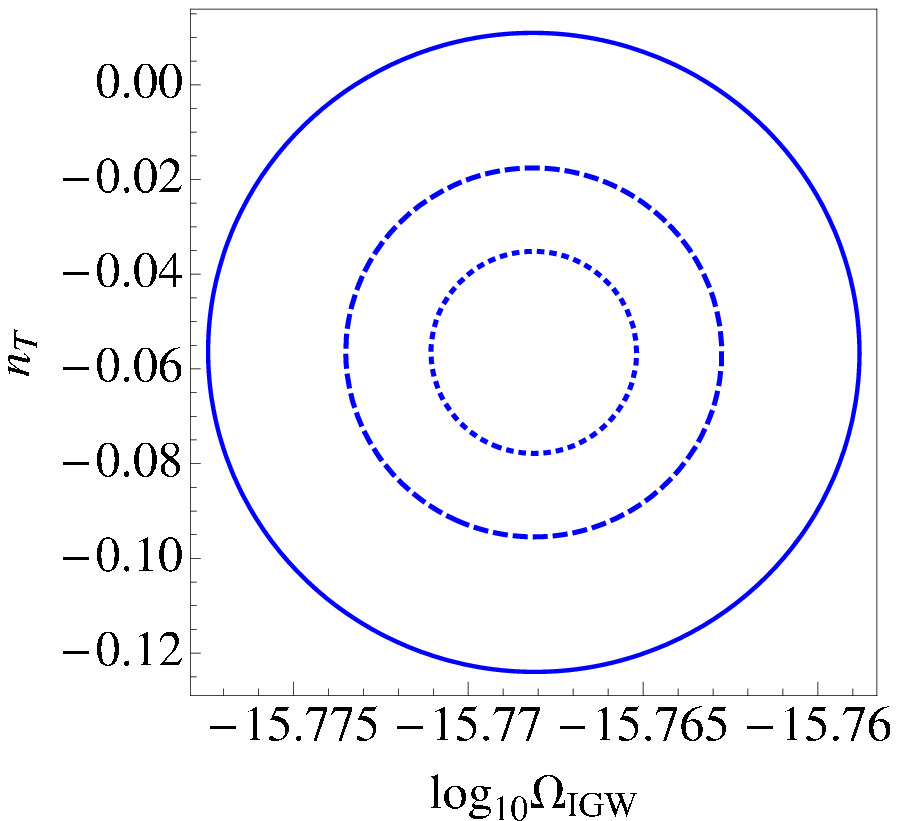}
    \end{center}
  \end{minipage}
  \begin{minipage}{0.4\columnwidth}
    \begin{center}
      \includegraphics[clip, width=1.0\columnwidth]{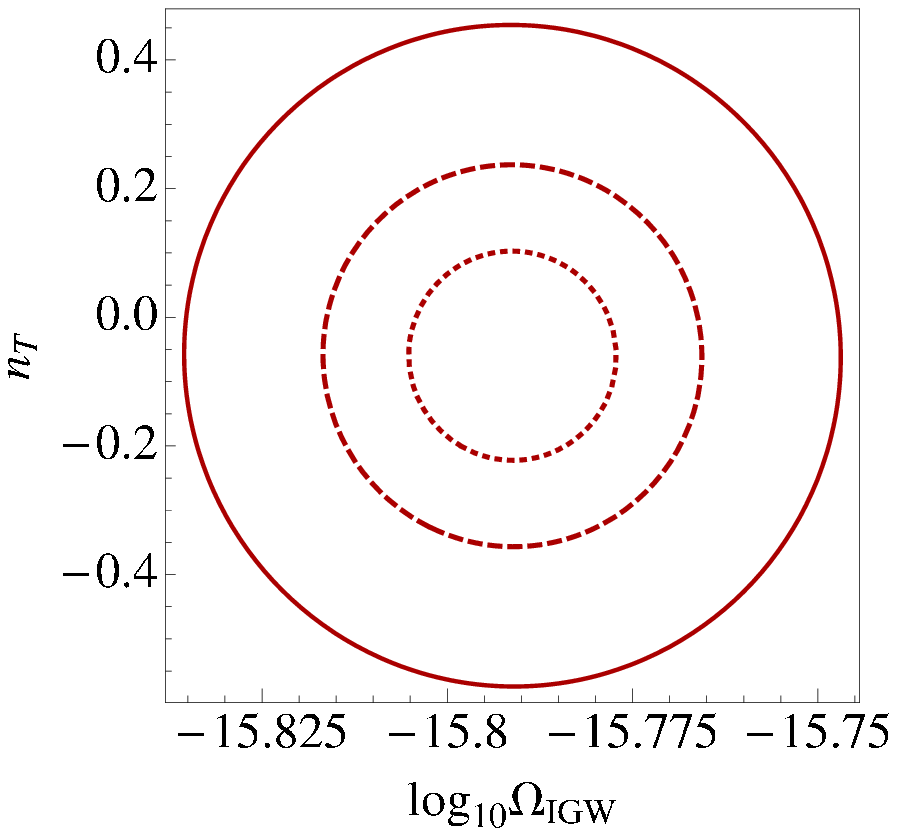}
    \end{center}
  \end{minipage}
 
  \begin{minipage}{0.4\columnwidth}
    \begin{center}
      \includegraphics[clip, width=1.0\columnwidth]{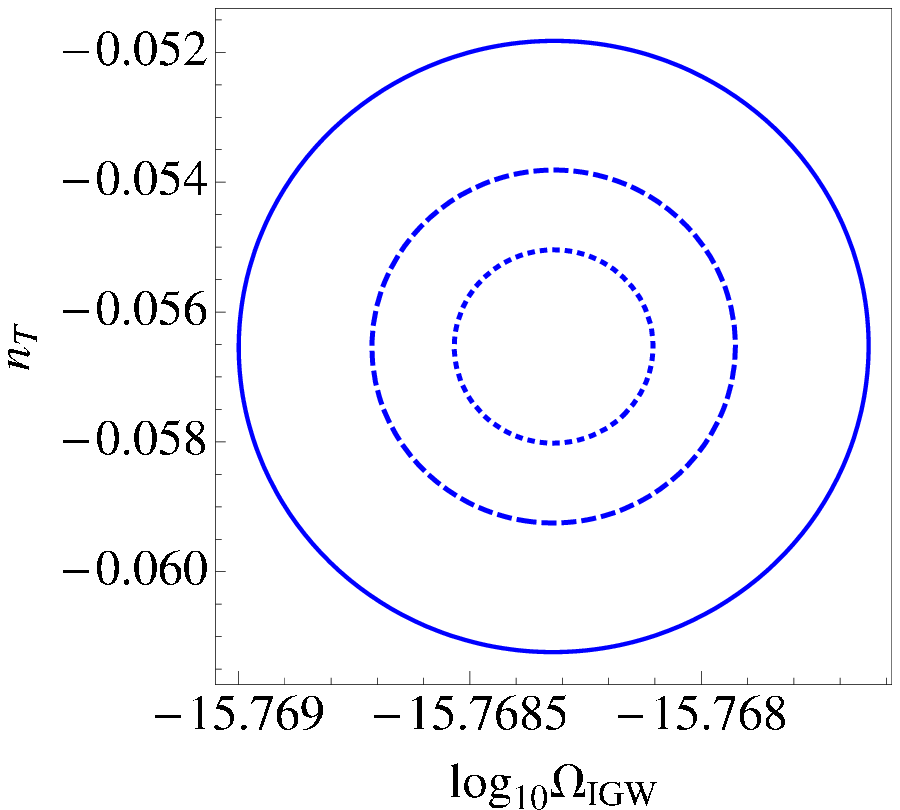}
    \end{center}
  \end{minipage}
  \begin{minipage}{0.4\columnwidth}
    \begin{center}
      \includegraphics[clip, width=1.0\columnwidth]{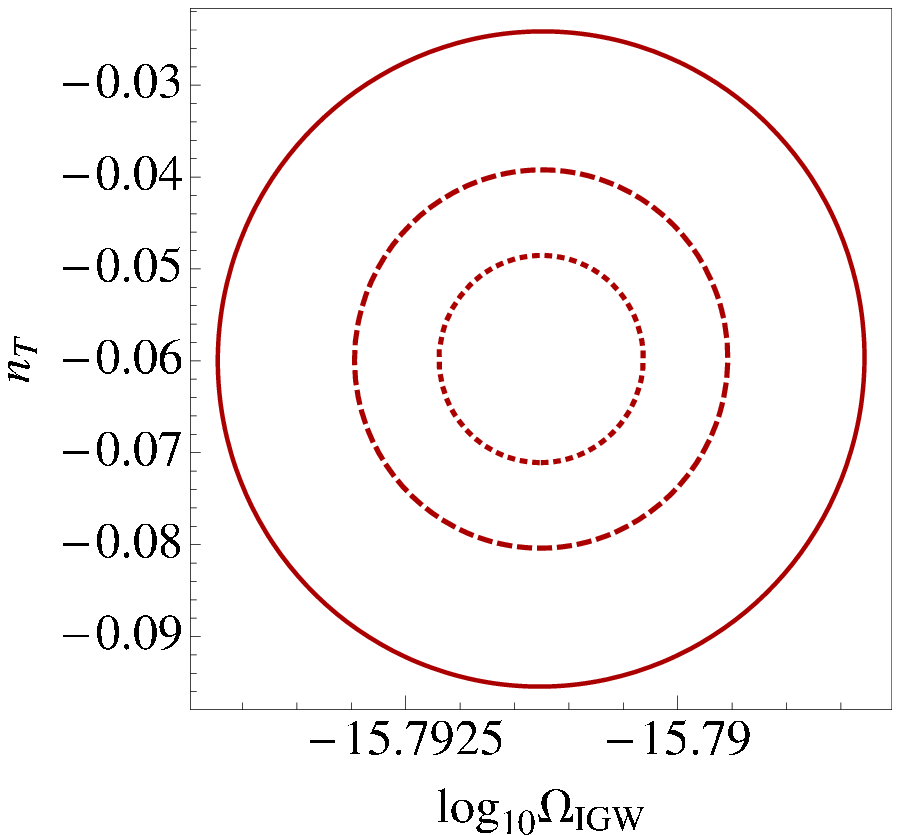}
    \end{center}
  \end{minipage}
  \end{center}
  \caption{\small $95\%$ C.L. expected constraints on the $\log_{10}
    \bar{\Omega}_{\rm IGW}$ vs.~$n_{\rm T}$ plane for BBO-std (top),
    BBO-grand (middle) and ult-DECIGO (bottom).  In these figures,
    $\log_{10} \bar{\Omega}_{\rm IGW}$ and $n_{\rm T}$ are varied while
    $\alpha_{\rm T}$ is fixed to the fiducial value.  The minimum
    frequency $f_{\rm min}$ is taken to be $0.1\ {\rm Hz}$ (left
    panels with blue contours) and $0.3\ {\rm Hz}$ (right panels with
    red contours).  The solid (dashed, dotted) line corresponds to
    $T_{\rm obs}=$1\ yr, (3\ yr, 10\ yr).}
  \label{fig:highTR_omegant}
\end{figure}

\begin{figure}
  \begin{center}
    \begin{minipage}{0.4\columnwidth}
      \begin{center}
        \includegraphics[clip, width=1.0\columnwidth]{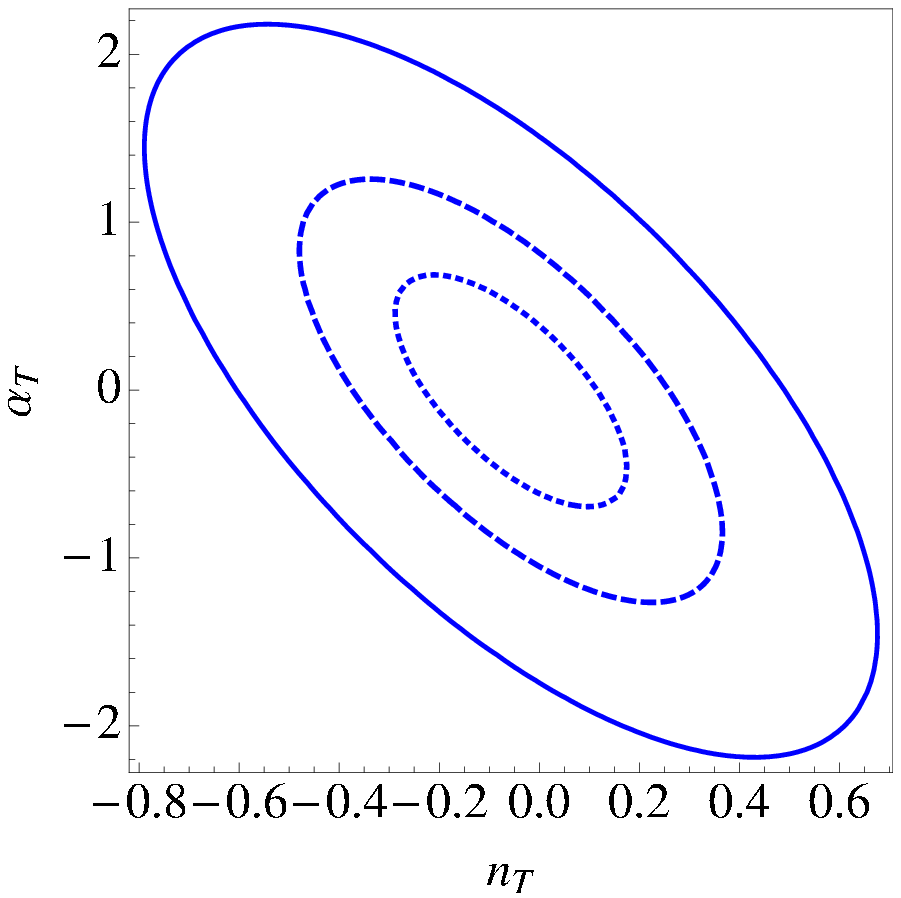}
      \end{center}
    \end{minipage}
    \begin{minipage}{0.4\columnwidth}
      \begin{center}
        \includegraphics[clip, width=1.0\columnwidth]{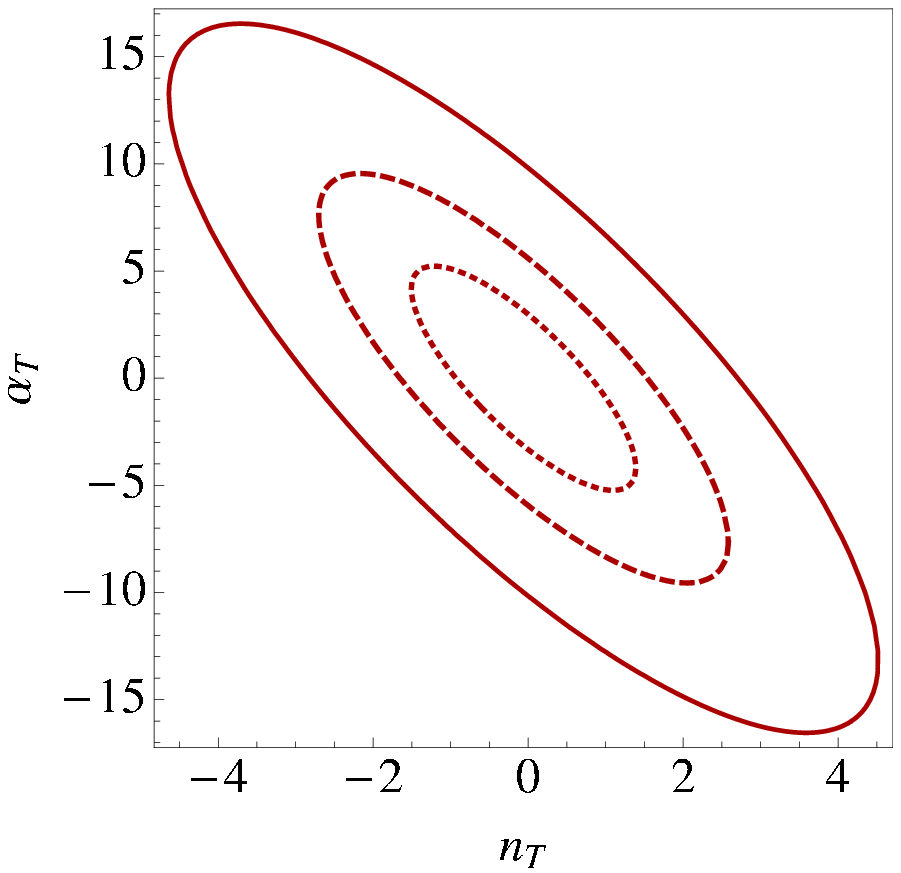}
      \end{center}
    \end{minipage}  

    \begin{minipage}{0.4\columnwidth}
      \begin{center}
        \includegraphics[clip, width=1.0\columnwidth]{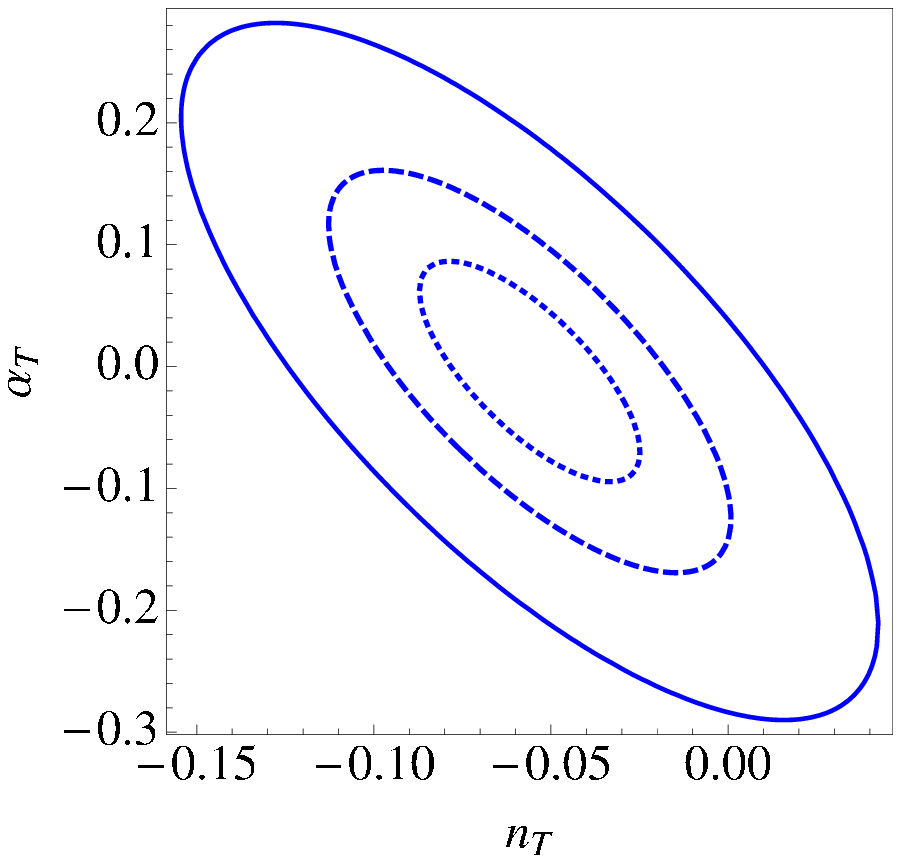}
      \end{center}
    \end{minipage}
    \begin{minipage}{0.4\columnwidth}
      \begin{center}
        \includegraphics[clip, width=1.0\columnwidth]{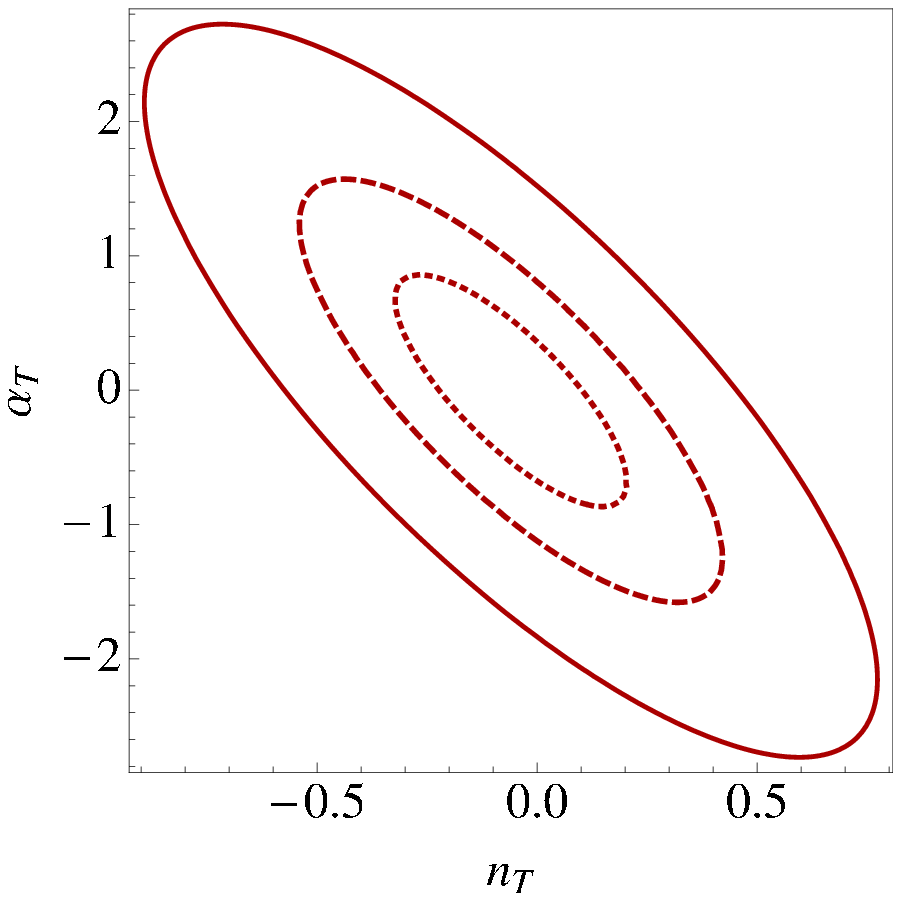}
      \end{center}
    \end{minipage} 

    \begin{minipage}{0.4\columnwidth}
      \begin{center}
        \includegraphics[clip, width=1.0\columnwidth]{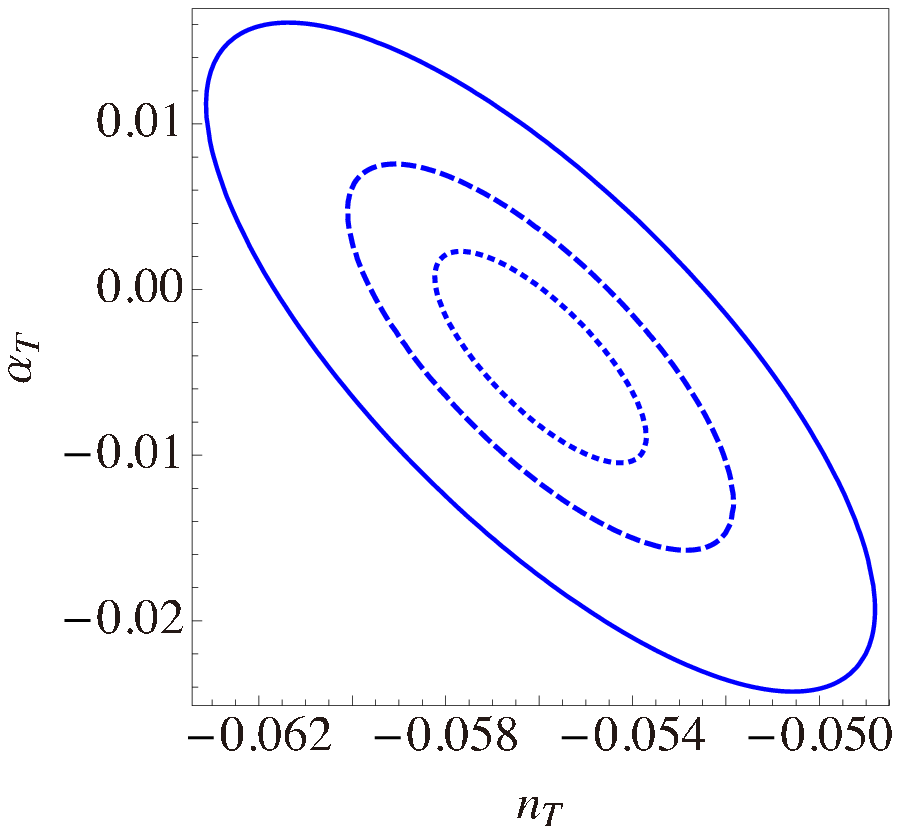}
      \end{center}
    \end{minipage}
    \begin{minipage}{0.4\columnwidth}
      \begin{center}
        \includegraphics[clip, width=1.0\columnwidth]{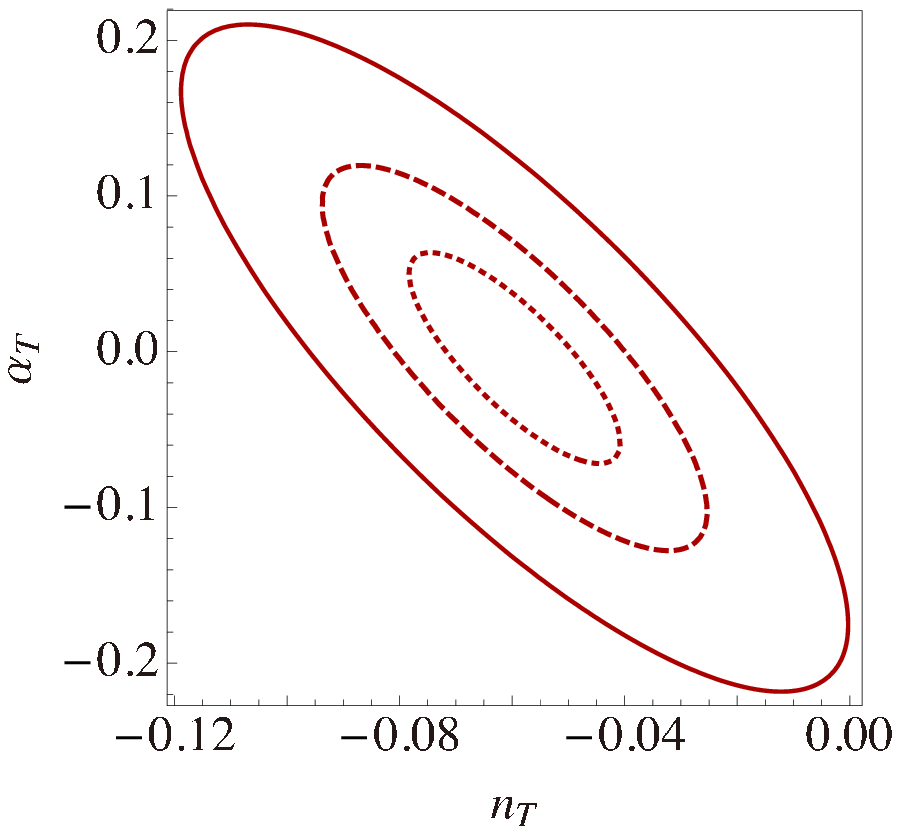}
      \end{center}
    \end{minipage}

    \end{center}
    
      \caption{\small $95\%$ C.L. expected constraints on the $n_{\rm T}$ vs.~$\alpha_{\rm T}$ plane for 
  BBO-std (top), BBO-grand (middle) and ult-DECIGO (bottom).
  In these figures, $\log_{10} \bar{\Omega}_{\rm IGW}$ is marginalized. 
   The minimum frequency $f_{\rm min}$ is taken to be $0.1\ {\rm Hz}$ (left panels with blue contours) and $0.3\ {\rm
      Hz}$ (right panels with red contours).  The solid (dashed, dotted) line corresponds to
    $T_{\rm obs}=$1\ yr, (3\ yr, 10\ yr).}
  \label{fig:highTR_ntalphat}
\end{figure}

We calculate the likelihood as a function of $\{ p\}=\{
\log_{10}\bar{\Omega}_{\rm IGW}, n_{\rm T}, \alpha_{\rm T} \}$.  In the left
panels of Figs.\ \ref{fig:highTR_omegant} $-$
\ref{fig:highTR_ntalphat}, we show the contours of constant
$\delta\chi^2=5.99$ (corresponding to 95~\% C.L.) on the
$\bar{\Omega}_{\rm IGW}$ vs.\ $n_{\rm T}$ and $n_{\rm T}$ vs.\
$\alpha_{\rm T}$ planes.\footnote
{Here and hereafter, axes of some of the panels are extended to
  very large values, like $n_{\rm T}\sim O(1)$ and $\alpha_{\rm T}\sim
  10$.  These are taken just for demonstrative purposes, although
  they are unnaturally large.}
In each figure, the noise level of BBO-std, BBO-grand, or ult-DECIGO
is adopted, and the lowest frequency $f_{\rm min}$ is taken to be
$0.1\ {\rm Hz}$. 

First, we consider the determination of $\bar{\Omega}_{\rm IGW}$ and
$n_{\rm T}$.  As we will see below, the uncertainty of $\alpha_{\rm
  T}$ becomes comparable to or larger than that of $n_{\rm T}$.  On
the contrary, in slow-roll inflation model, $|\alpha_{\rm T}|$ is
expected to be much smaller than $|n_{\rm T}|$.  This implies that,
assuming slow-roll inflation, the running is (almost) irrelevant for
the determination of $\bar{\Omega}_{\rm IGW}$ and $n_{\rm T}$.  Thus,
we first discuss the result based on the analysis with $\{ p \}=\{
\log_{10}\bar{\Omega}_{\rm IGW}, n_{\rm T} \}$ and $\alpha_{\rm T}$ being
fixed to the fiducial value.  Here we note that $\{p\}$ depends on the
pivot scale $f_*$, i.e. $\{p\} = \{p(f_*)\}$.  We can make the
correlation between $\ln \Omega_{\rm IGW}$ and $n_{\rm T}$ vanish by
properly choosing $f_*$; for such a choice of $f_*$, the error of
$\bar{\Omega}_{\rm IGW}$ is minimized if we neglect $\alpha_{\rm
  T}$.\footnote
{For notational simplicity, let us denote $x_0=\ln \bar{\Omega}_{\rm
    IGW}- \ln \hat{\bar{\Omega}}_{\rm IGW}$ and $x_1=n_{\rm
    T}-\hat{n}_{\rm T}$.  Then, neglecting $\alpha_{\rm T}$, the
  amplitude and the tensor spectral index for two different pivot
  scales, $f_*$ and $f_*'$, are related as
\begin{align*}
  x_0'
  = x_0 + x_1 \ln (f_*' / f_*), ~~~
  x_1'
  = x_1,
\end{align*}
where the quantities without (with) the prime are evaluated at
$f_*$ ($f'_*$).  (See Eqs.\ \eqref{Omega'} and \eqref{nT'}.)  The
error of $x_0'$ is given by
\begin{align*}
  \langle x_0'^2 \rangle
  = \langle x_0^2 \rangle + 2\ln (f_*' / f_*) \langle x_0 x_1 \rangle
  + \ln^2 (f_*' / f_*) \langle x_1^2 \rangle,
\end{align*}
where $\langle\cdots\rangle$ denotes the expectation value calculated
with Gaussian probability density.  Then, $\langle x_0'^2 \rangle$ is
minimized when
\begin{align*}
  \ln (f_*' / f_*)
  = - \langle x_0 x_1 \rangle /  \langle x_1^2 \rangle.
\end{align*}
With such a choice of $f'_*$, $\langle x_0' x_1'\rangle=0$, and hence
$x_0'$ and $x_1'$ become uncorrelated.  
}
Thus in the following we choose such $f_*$ for each experiment and
each $f_{\rm min}$, and the value is summarized in Table
\ref{table:fstar}.

On $\bar{\Omega}_{\rm IGW}$, a relatively good determination of
$\bar{\Omega}_{\rm IGW}$ is possible even with the noise level of
BBO-std \cite{Seto:2005qy}. (See the top panels of Fig.\
\ref{fig:highTR_omegant}.)  With a better noise level, like BBO-grand
and ult-DECIGO, more precise measurement of $\bar{\Omega}_{\rm IGW}$
is expected.  (See the middle and bottom of the same figure.)  From
the Fisher matrix, we estimate the error in the determination of
$\bar{\Omega}_{\rm IGW}$; taking $T_{\rm obs}=10$ yr, the 1$\sigma$
error is found to be $2.6\ \%$, $0.28\ \%$, and $0.020\ \%$ for
BBO-std, BBO-grand, and ult-DECIGO, respectively, after marginalizing
$n_{\rm T}$.  The errors with other choices of fundamental parameters
are summarized in Table \ref{table:1sigma}.

\begin{table}[t]
  \begin{center} 
   \begin{tabular}{l|ccc}
      \hline\hline
      {} & {BBO-std 0.1/0.3Hz} 
      & {BBO-grand 0.1/0.3Hz} 
      & {ult.-DECIGO 0.1/0.3Hz}
      \\
      \hline
      {$\bar{\Omega}_{\rm IGW},\; n_{\rm T}$}
      & {$0.20$}/{$0.37$}
      & {$0.16$}/{$0.37$}
      & {$0.17$}/{$0.37$}\\
      \hline
      {$\bar{\Omega}_{\rm IGW},\; n_{\rm T},\; \alpha_{\rm T}$}
      & {$0.17$}/{$0.35$}
      & {$0.15$}/{$0.35$}
      & {$0.14$}/{$0.35$}\\
   \hline\hline
    \end{tabular}
    \caption{$f_*$[Hz] for each experiment and for each
      cutoff frequency.}
    \label{table:fstar}

\vspace{1em}

    \begin{tabular}{l|ccc}
      \hline\hline
      {} & {BBO-std 0.1/0.3Hz} 
      & {BBO-grand 0.1/0.3Hz} 
      & {ult.-DECIGO 0.1/0.3Hz}
      \\
      \hline
      {S/N}
      & {$39$}/{$14$}
      & {$3.6\times 10^2$}/{$76$}
      & {$4.9\times 10^3$}/{$1.1 \times 10^3$}\\
      \hline
      {$\log_{10} \bar{\Omega}_{\rm IGW}$} 
      & {$1.1\times 10^{-2}$}/{$3.1\times 10^{-2}$}
      & {$1.2\times 10^{-3}$}/{$5.7\times 10^{-3}$}
      & {$8.8\times 10^{-5}$}/{$3.8\times 10^{-4}$}\\
      {$\log_{10} \bar{\Omega}_{\rm IGW}$ ${\scriptstyle {\rm (w/}n_{\rm T}{\rm )}}$}
      & {$1.4\times 10^{-2}$}/{$3.3\times 10^{-2}$}
      & {$1.4\times 10^{-3}$}/{$6.1\times 10^{-3}$}
      & {$1.0\times 10^{-4}$}/{$4.2\times 10^{-4}$}\\
      {$\log_{10} \bar{\Omega}_{\rm IGW}$ ${\scriptstyle {\rm (w/}n_{\rm T},\alpha_{\rm T}{\rm )}}$} 
      & {$1.3\times 10^{-2}$}/{$3.3\times 10^{-2}$}
      & {$1.4\times 10^{-3}$}/{$6.1\times 10^{-3}$}
      & {$1.0\times 10^{-4}$}/{$4.1\times 10^{-4}$}\\
      \hline
      {$n_{\rm T}$ ${\scriptstyle {\rm (w/}\log_{10} \bar{\Omega}_{\rm IGW}{\rm )}}$} 
      & {$7.1\times 10^{-2}$}/{$0.36$}
      & {$8.7\times 10^{-3}$}/{$6.6\times 10^{-2}$}
      & {$6.1\times 10^{-4}$}/{$4.6\times 10^{-3}$}\\
      {$n_{\rm T}$ ${\scriptstyle {\rm (w/}\log_{10} \bar{\Omega}_{\rm IGW},\alpha_{\rm T}{\rm )}}$}
      & {$9.5\times 10^{-2}$}/{$0.59$}
      & {$1.3\times 10^{-2}$}/{$0.11$}
      & {$9.3\times 10^{-4}$}/{$7.7\times 10^{-3}$}\\
      \hline
      {$\alpha_{\rm T}$ ${\scriptstyle {\rm (w/}\log_{10} \bar{\Omega}_{\rm IGW},n_{\rm T}{\rm )}}$}
      & {$0.28$}/{$2.1$}
      & {$3.7\times 10^{-2}$}/{$0.35$}
      & {$2.6\times 10^{-3}$}/{$2.8\times 10^{-2}$}\\
      \hline\hline
    \end{tabular}
    \caption{1$\sigma$ error for each parameter with the lowest
      frequency $f_{\rm min}=0.1$ and $0.3\ {\rm Hz}$.  Here, we
      take $T_{\rm obs}=10\ {\rm yr}$. (Notice that the errors scale as
      $T_{\rm obs}^{-1/2}$.) In the parenthesis,
    the parameters marginalized are listed. }
    \label{table:1sigma}  
    \end{center}
\end{table}

The error of $n_{\rm T}$ can be also understood from Fig.\
\ref{fig:highTR_omegant}.  We also calculate the error of $n_{\rm T}$
for each detector parameters; the results are summarized in Table
\ref{table:1sigma}.  We can see that, even with the noise level of
BBO-std, $n_{\rm T}$ may be known to be $O(0.01)$.  If such a result
becomes available, it will tell us that the slow-roll condition is
likely to be satisfied when the mode with $f\sim 0.1\ {\rm Hz}$ exits
the horizon during inflation.  It is notable that, with the noise
level of BBO-grand, the uncertainty of $n_{\rm T}$ becomes comparable
to the fiducial value (if we adopt the prediction of the chaotic
inflation).  This fact implies that, with such a sensitivity, we may
be able to detect the tensor spectral index which provides a very
important information about the slow-roll parameter.  Then, with
ult-DECIGO, the error of $n_{\rm T}$ can be $\lesssim 10^{-3}$.

In order to estimate the expected sensitivity of the running (as well
as others), we also calculate the likelihood taking $\{ p
\}=\{ \log_{10}\bar{\Omega}_{\rm IGW}, n_{\rm T}, \alpha_{\rm T} \}$.  The
result is shown in Fig.\ \ref{fig:highTR_ntalphat}.  In making these
figures we chose $f_*$ for which the correlation between
$\bar{\Omega}_{\rm IGW}$ and $n_{\rm T}$ vanishes after marginalizing
$\alpha_{\rm T}$.  Here we show only the $n_{\rm T}$ vs.\ $\alpha_{\rm T}$
plane after marginalizing $\log_{10}\bar{\Omega}_{\rm IGW}$, since
the $\bar{\Omega}_{\rm IGW}$ vs.\ $n_{\rm T}$ plane with $\alpha_{\rm T}$
being marginalized is roughly the same as Fig.\ \ref{fig:highTR_omegant}
(about $O(10\%)$ difference in the error of each parameter).  We can
see that, with the noise level of BBO-std, the error in $\alpha_{\rm
  T}$ is orders of magnitude larger than the expectation from
slow-roll inflation.  With the noise level of BBO-grand, on the
contrary, we may obtain a bound on $|\alpha_{\rm T}|$ of $\lesssim
0.1$ if the operation time of $\sim 10\ {\rm yr}$ is adopted.  If the
noise level of ult-DECIGO is available, we have a stronger bound of
$\lesssim 0.01$.

So far, we have shown the results for $f_{\rm min}=0.1\ {\rm Hz}$.
However, we should note here that the accuracy of the parameter
determination strongly depends on the lowest frequency $f_{\rm min}$.
To see what happens if we take larger value of $f_{\rm min}$, we also
calculate the likelihood with $f_{\rm min}=0.3\ {\rm Hz}$.
The results for BBO-std, BBO-grand and ult-DECIGO are shown in the
right panels of Figs.\ \ref{fig:highTR_omegant} $-$
\ref{fig:highTR_ntalphat}. As we can see, larger value of $f_{\rm min}$
results in a worse measurement of the fundamental parameters.  This is
because the signal-to-noise ratio becomes largest for $f\sim O(0.1\
{\rm Hz})$.  As we have mentioned, the value of $f_{\rm min}$ reflects
the expectation that the density of the GWs from white-dwarf binaries
is much larger than that of IGWs.  Better understanding of the former
would help to improve the study of the IGWs.

\begin{figure}
  \begin{center}
    \begin{minipage}{0.5\columnwidth}
      \begin{center}
        \includegraphics[clip, width=1.0\columnwidth]{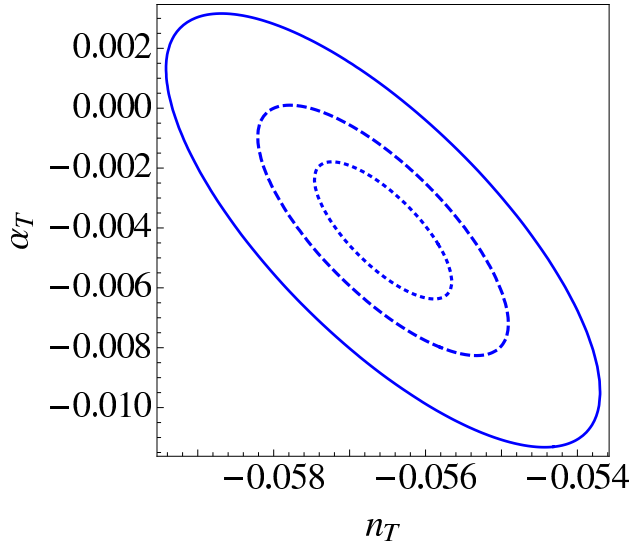}
      \end{center}
    \end{minipage}
    \end{center}
    \caption{\small 95\ \% C.L. expected constraints on the
$n_{\rm T}$ vs.~$\alpha_{\rm T}$ plane for the 
      detector parameters given in Eqs.\ (\ref{eq:superult_shot}) and
      (\ref{eq:superult_accel}) with $f_{\rm min}=0.1\ {\rm Hz}$.  
The solid (dashed, dotted) line corresponds to
    $T_{\rm obs}=$1\ yr, (3\ yr, 10\ yr).
The
      pivot scale is taken to be $f_* = 0.17$ Hz.}
  \label{fig:highTR_ntalphat_superult}
\end{figure}

To distinguish $\alpha_{\rm T}$ from $0$, we need a more sensitive
detector than ult-DECIGO. Since the standard quantum limit depends on
$L$ and $m$, we consider the case with $L=5\times 10^8$ km and
$m=500$ kg, for example.  Then, assuming that the noise level is
limited only by the standard quantum limit and that the shot noise
dominates the frequency range of $f>0.1\ {\rm Hz}$, we take
\begin{align}
S_y^{\rm optical-path} 
&= 9.3 \times 10^{-52} \; [(f/{\rm Hz})^2 \; {\rm Hz}^{-1}],
\label{eq:superult_shot} \\
S_y^{\rm proof-mass}
&= 9.3 \times 10^{-56} \; [(f/{\rm Hz})^{-2} \; {\rm Hz}^{-1}].
\label{eq:superult_accel}
\end{align}
The expected constraint on the $n_{\rm T}$ vs.\ $\alpha_{\rm T}$ plane (with
$f_{\rm min}=0.1$ Hz and $f_*=0.17$ Hz) is shown in Fig.\
\ref{fig:highTR_ntalphat_superult}.  With such a noise level,
non-vanishing value of $\alpha_{\rm T}$ may be seen; then, combining
the information about $\alpha_{\rm T}$ and $n_{\rm T}$, the slow-roll
parameters $\epsilon$ and $\eta$ can be reconstructed, which would
become a very important discriminator of various inflation models.

\subsection{Determination of $T_{\rm R}$}
\label{sec:highTR}

Next, we include $T_{\rm R}$ into the fit parameters, paying
particular attention to the determination of $T_{\rm R}$.  As we have
mentioned, the IGW spectrum for $f\gtrsim f_{\rm R}$ is significantly
suppressed.  If such a behavior can be confirmed by GW detectors, we
have a possibility to acquire the information about the reheating
temperature \cite{Nakayama:2008ip,Nakayama:2008wy,Kuroyanagi:2009br,
  Nakayama:2009ce,Kuroyanagi:2011fy,Kuroyanagi:2013ns}.  It should be,
however, also noted that the change of the shape of the IGW spectrum
may affect the determinations of other fundamental parameters, in
particular, $\bar{\Omega}_{\rm IGW}$ and $n_{\rm T}$.  In this
subsection, we consider how well we can determine the shape of the
IGWs, taking $\bar{\Omega}_{\rm IGW}$, $n_{\rm T}$, and $T_{\rm R}$ as
fundamental parameters.  In this analysis, we do not include the
parameter $\alpha_{\rm T}$, because $|\alpha_{\rm T}|$ is
expected to be small in slow-roll inflation model.  Also, we choose
the pivot scale $f_*$ to be the same value as adopted in the analysis
with $\bar{\Omega}_{\rm IGW}$ and $n_{\rm T}$ in the previous
subsection (see Table \ref{table:fstar}).

\begin{figure}[t]
  \centerline{\epsfxsize=0.6\textwidth\epsfbox{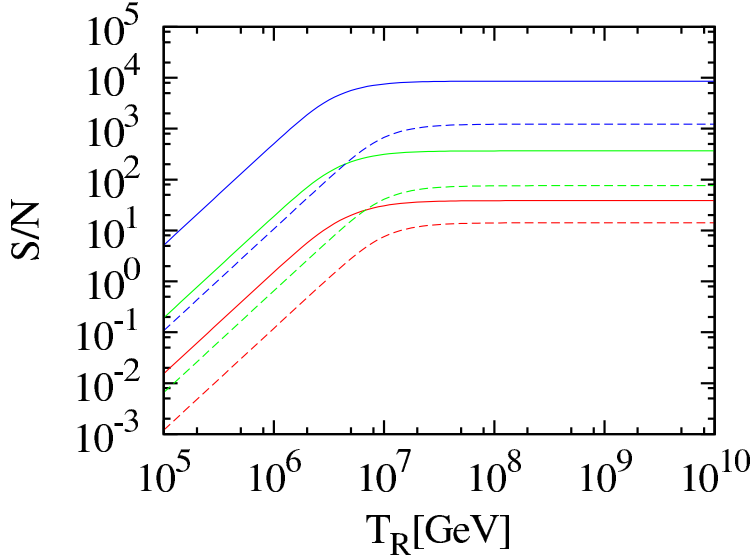}}
  \caption{\small Signal-to-noise ratio as a function of the
    reheating temperature.  The IGW spectrum is calculated with
    $\bar{\Omega}_{\rm IGW}=1.51\times 10^{-16}$, and $n_{\rm
      T}=-6.38\times 10^{-2}$. Each line corresponds to BBO-std (red),
    BBO-grand (green) and ult-DECIGO (blue), respectively, and the
    solid and dashed lines are for $f_{\rm min}=0.1$ and for $0.3$
    Hz, respectively. The observation time is taken to be $T_{\rm
      obs}=10\ {\rm yr}$. }
  \label{fig:snTR}
\end{figure}

If the reheating temperature is relatively low, the IGW amplitude in
the frequency range relevant for the GW detectors is suppressed, as we
have discussed in Section \ref{sec:spectrum}.  Then, with low
reheating temperature, the detection of the IGW signal becomes
difficult.  To see this, in Fig.\ \ref{fig:snTR}, we show the
signal-to-noise ratio as a function of $T_{\rm R}$.  As one can see, the
detection of the IGW spectrum is possible only if $T_{\rm R}\gtrsim
10^5-10^6\ {\rm GeV}$.  Thus, in the following, we concentrate on the
case where the reheating temperature is higher than $10^6\ {\rm GeV}$.

\begin{figure}
  \centerline{\epsfysize=0.8\textheight\epsfbox{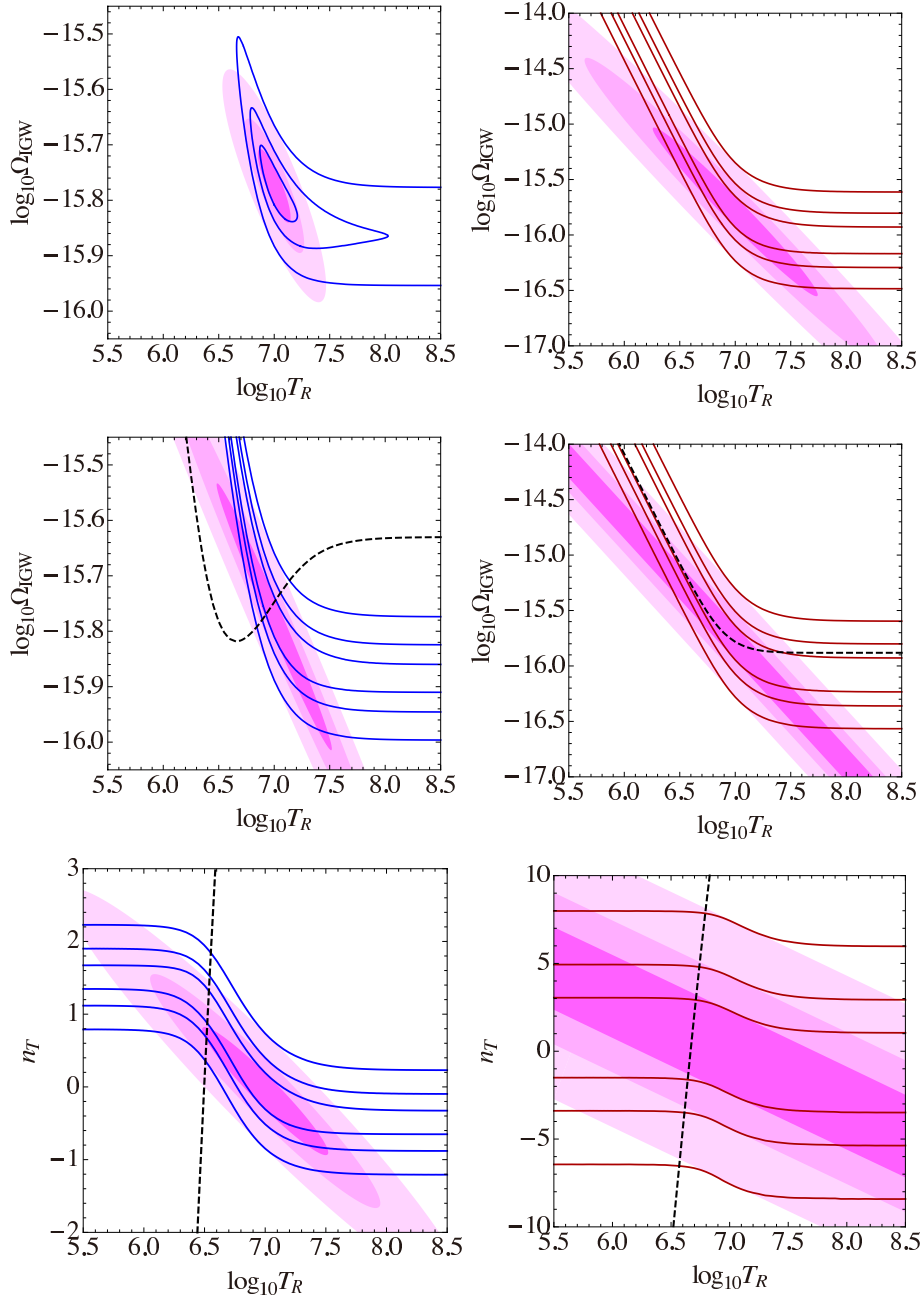}}
  \caption{\small The solid lines are the contours of
    $\delta\chi^2=5.99$ with the fiducial values of
    $T_{\rm R}=10^7\ {\rm GeV}$.  The noise function of BBO-std is
    adopted with $T_{\rm obs}=1$, $3$, and $10\ {\rm yr}$ (from
    outside to inside), and $f_{\rm min}=0.1$ (left panels with blue
    lines) and $0.3\ {\rm Hz}$ (right panels with red lines).  Pink
    regions are the results in the case where $T_{\rm R}$ is also
    included in the Fisher analysis.  Top: case with $n_{\rm T}$ being
    fixed to be the fiducial value.  Middle: case with $n_{\rm T}$
    being marginalized.  The dashed line corresponds to the contour on
    which the best-fit value of $n_{\rm T}$ becomes $0$.  Bottom: case
    with $\bar{\Omega}_{\rm IGW}$ being marginalized.  The dashed line
    corresponds to the contour on which the best-fit value of
    $\bar{\Omega}_{\rm IGW}$ becomes $5\times 10^{-16}$.}
  \label{fig:lowTR_BBOstd_TR=10^7}
\end{figure}

\begin{figure}
  \centerline{\epsfysize=0.8\textheight\epsfbox{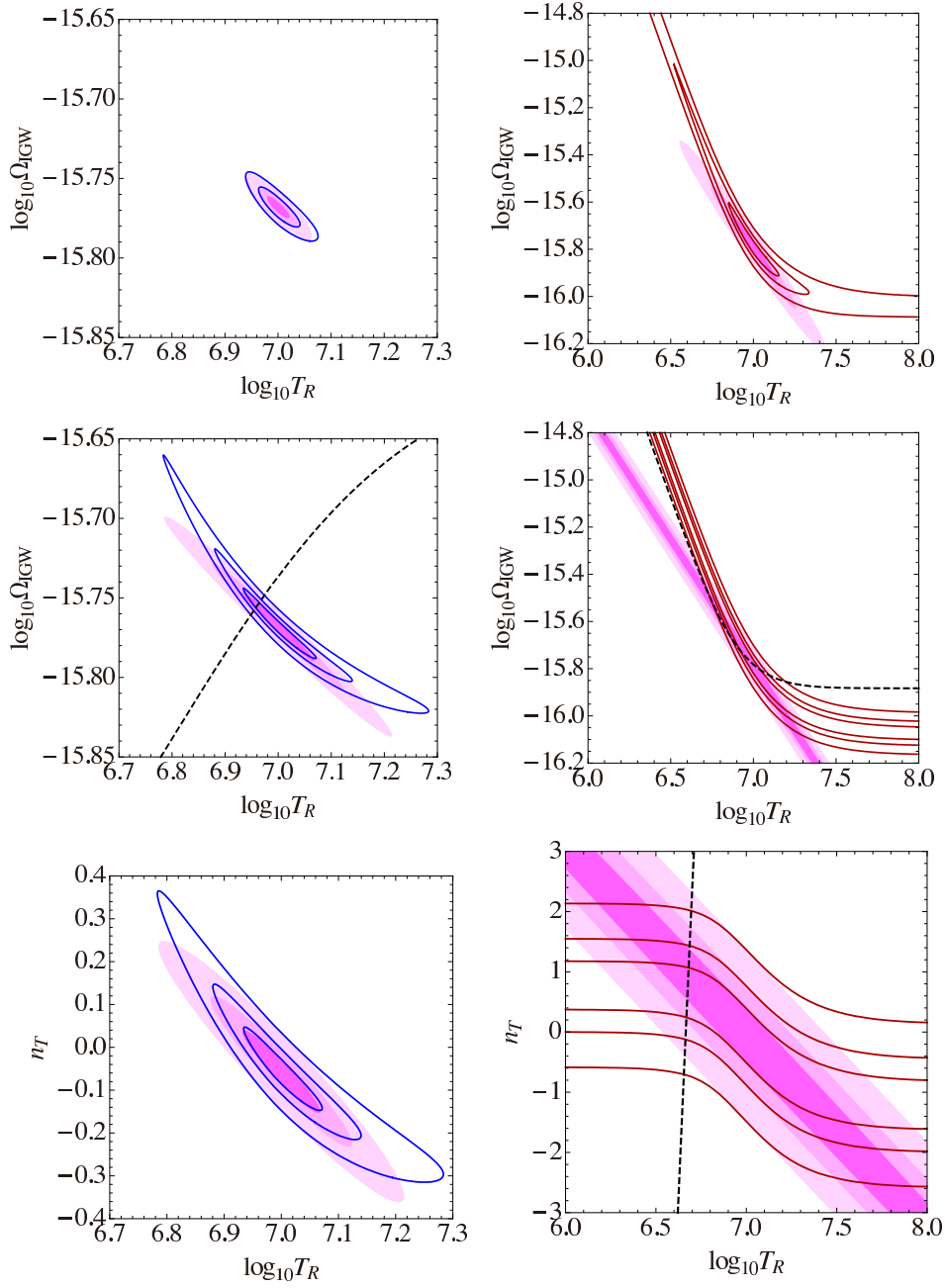}}
  \caption{\small Same as Fig.\ \ref{fig:lowTR_BBOstd_TR=10^7}, except that
   the noise function for BBO-grand is used.}
 \label{fig:lowTR_BBOgrand_TR=10^7}
\end{figure}

\begin{figure}
  \centerline{\epsfysize=0.8\textheight\epsfbox{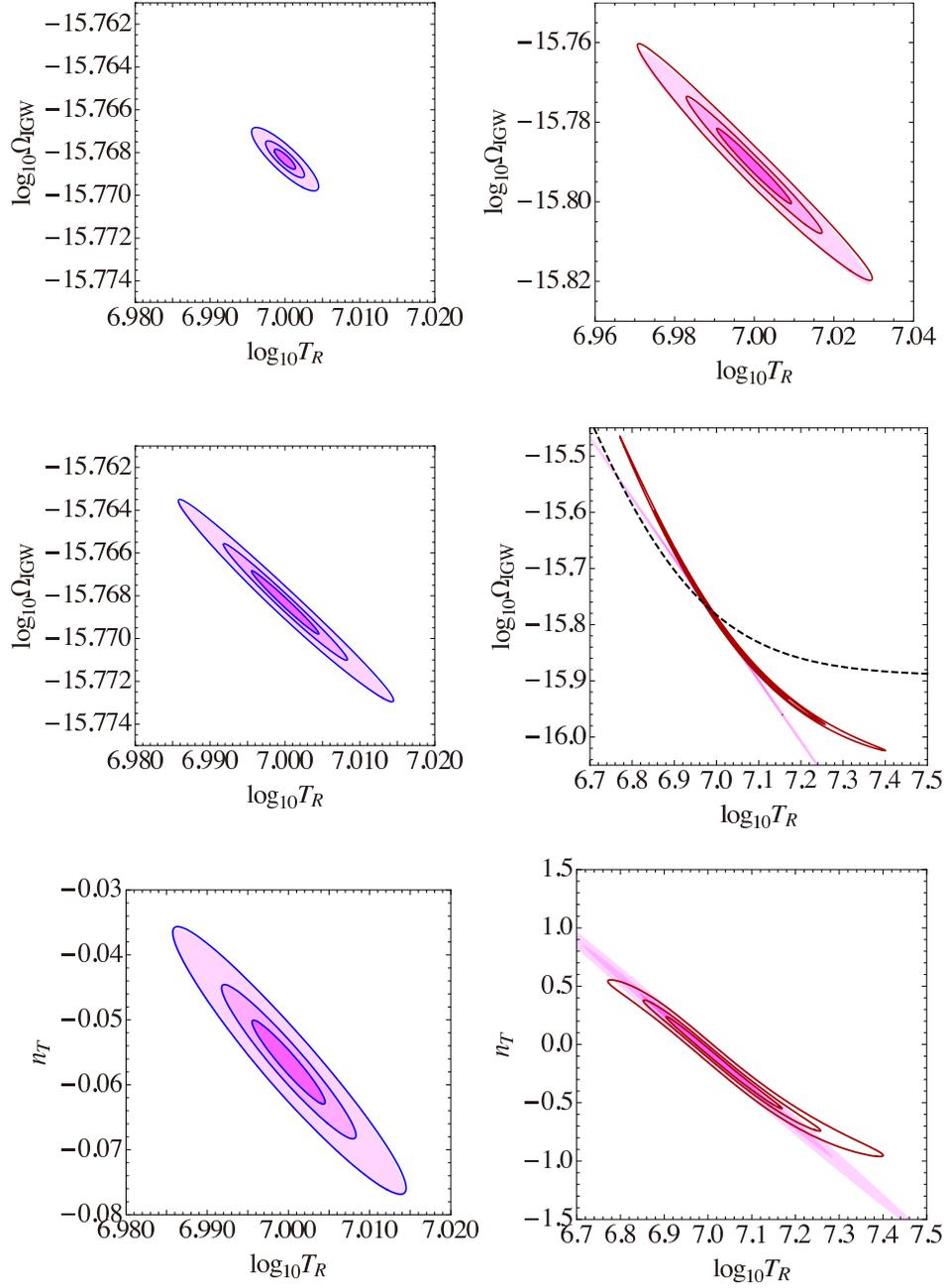}}
  \caption{\small Same as Fig.\ \ref{fig:lowTR_BBOstd_TR=10^7}, except
    that the noise function for ult-DECIGO is used.}
  \label{fig:lowTR_ulDECIGO_TR=10^7}
\end{figure}

\begin{figure}
  \centerline{\epsfysize=0.8\textheight\epsfbox{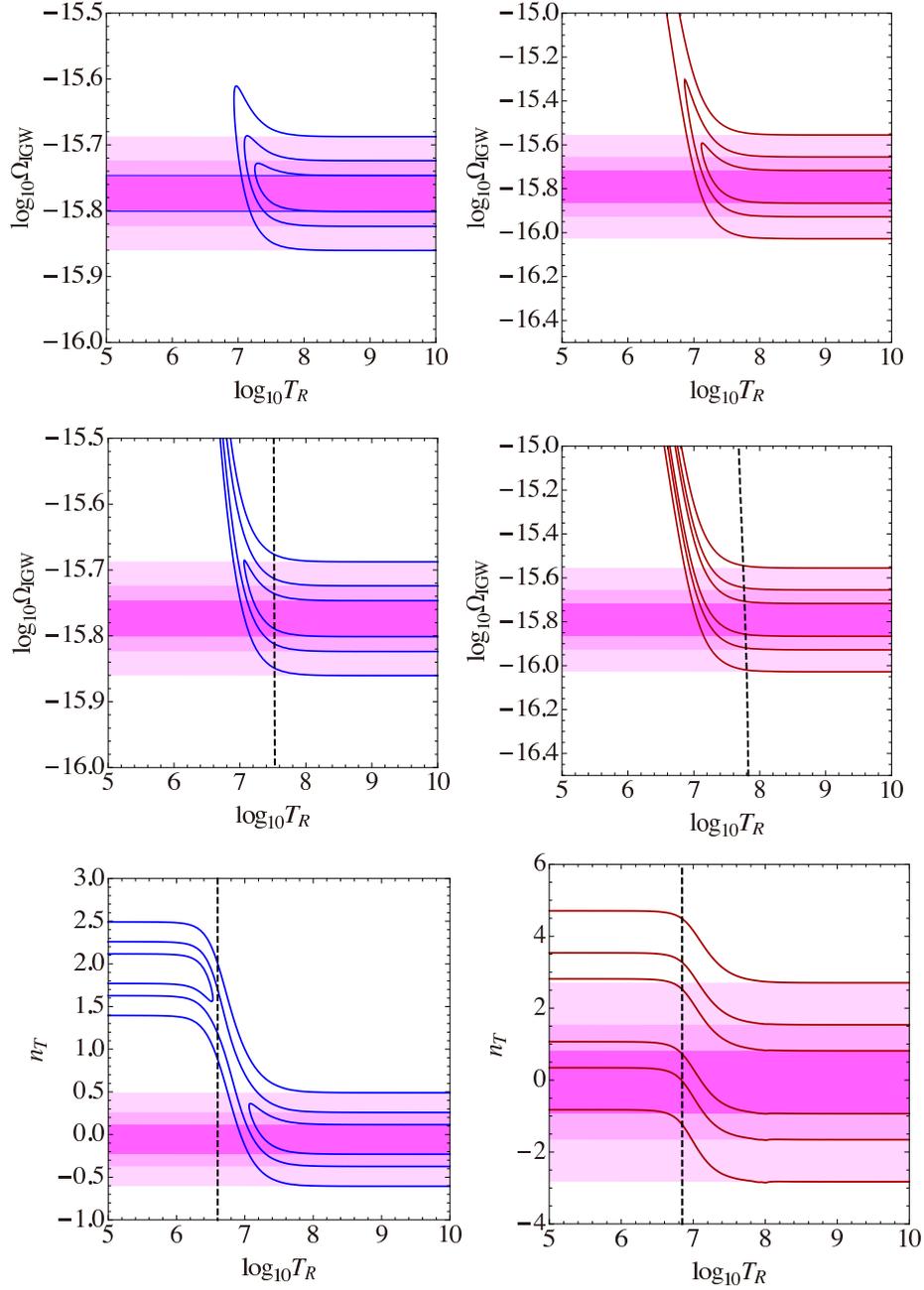}}
  \caption{\small Same as Fig.\ \ref{fig:lowTR_BBOstd_TR=10^7}, except
    for $T_{\rm R}=10^9 {\rm GeV}$.}
  \label{fig:lowTR_BBOstd_TR=10^9}
\end{figure}

\begin{figure}
  \centerline{\epsfysize=0.8\textheight\epsfbox{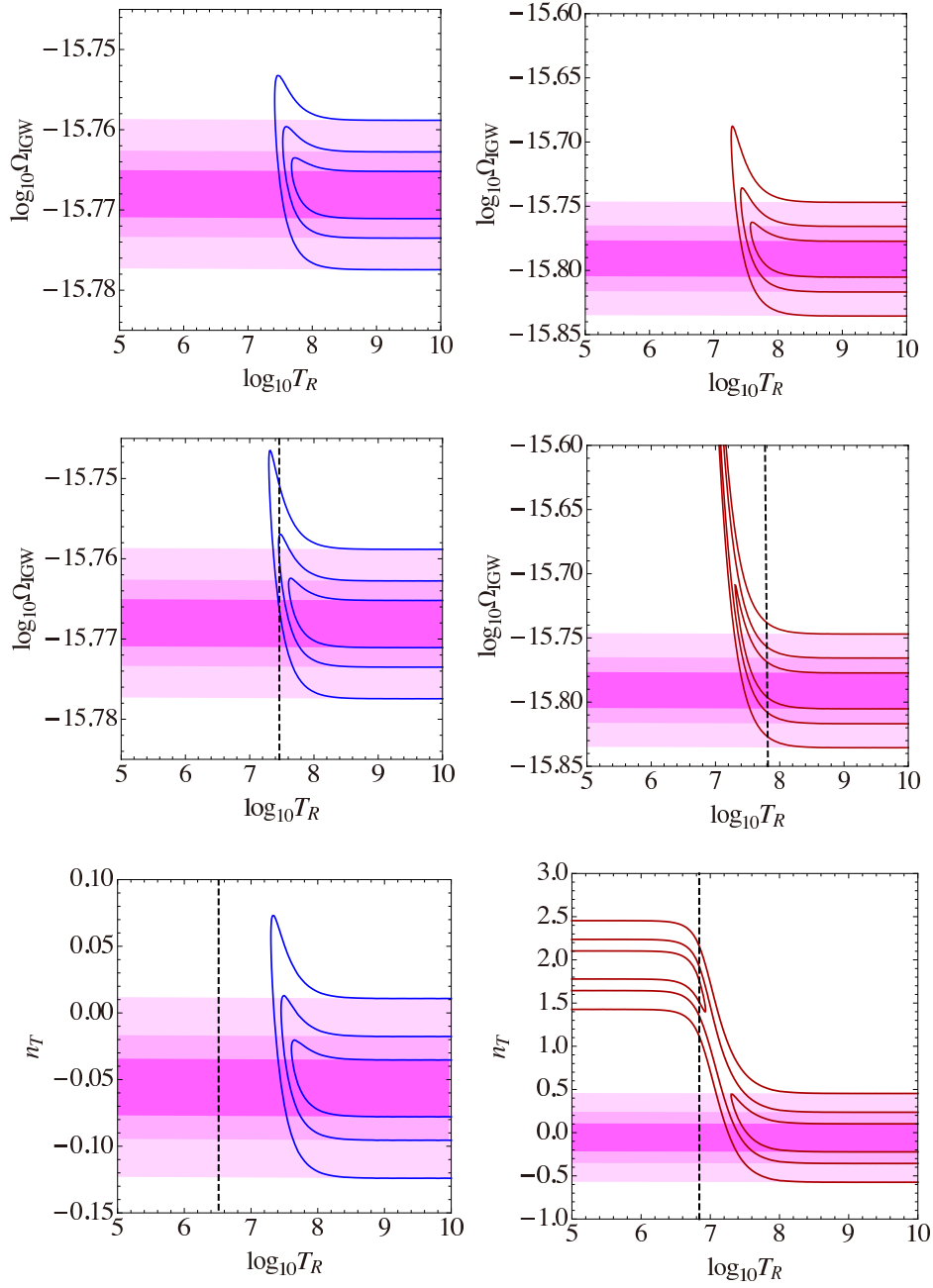}}
  \caption{\small Same as Fig.\ \ref{fig:lowTR_BBOstd_TR=10^7}, except 
    for $T_{\rm R}=10^9 {\rm GeV}$ and that the noise function for 
    BBO-grand is used.}
  \label{fig:lowTR_BBOgrand_TR=10^9}
\end{figure}

\begin{figure}
  \centerline{\epsfysize=0.8\textheight\epsfbox{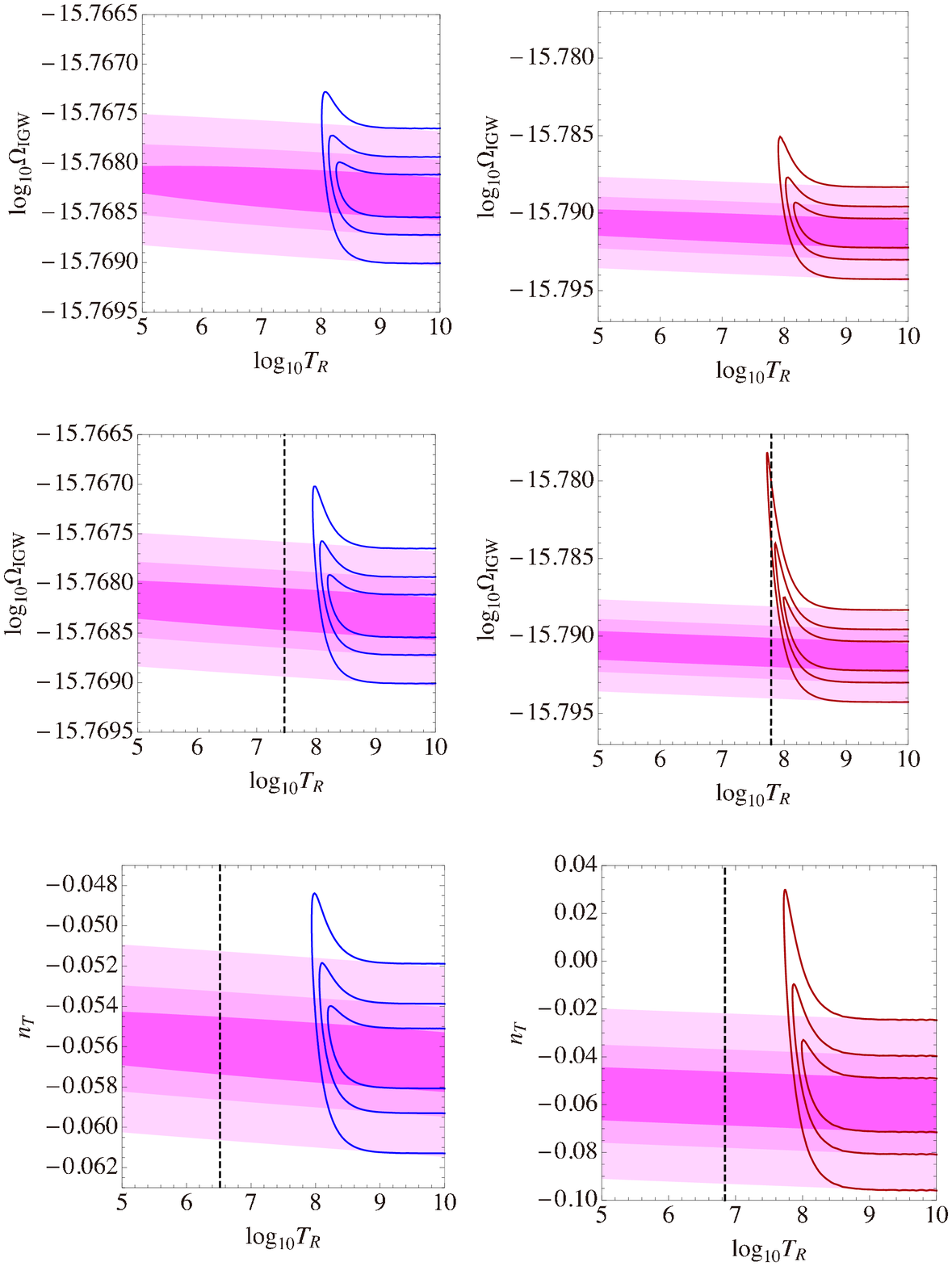}}
  \caption{\small Same as Fig.\ \ref{fig:lowTR_BBOstd_TR=10^7}, except
    for $T_{\rm R}=10^9 {\rm GeV}$ and that the noise function for
    ult-DECIGO is used.}
  \label{fig:lowTR_ulDECIGO_TR=10^9}
\end{figure}

We calculated the likelihood as a function of
$\bar{\Omega}_{\rm IGW}$, $n_{\rm T}$, and $T_{\rm R}$ for the case
where the signal-to-noise ratio is large enough for the detection.   Here,
$p_i^{\rm (ref)}$ is taken to be the point where ${\cal L}$ is
minimized for the fixed value of $T_{\rm R}$. In
Figs.\ \ref{fig:lowTR_BBOstd_TR=10^7} $-$
\ref{fig:lowTR_ulDECIGO_TR=10^9}, we show the contours of 
$\delta \chi^2 = 5.99$ on the $T_{\rm R}$ vs.\ $\bar{\Omega}_{\rm IGW}$ and $T_{\rm
  R}$ vs.\ $n_{\rm T}$ planes, taking the fiducial value of the
reheating temperature to be $10^7\ {\rm GeV}$ or $10^9\ {\rm
  GeV}$.\footnote
{In some figures, like Figs.\ \ref{fig:lowTR_BBOgrand_TR=10^7} 
and \ref{fig:lowTR_ulDECIGO_TR=10^7}, the
  allowed region shows island-like behavior. 
  They are the consequence of the poor data sampling in
  our numerical calculation due to our limitation of the computational
  power.}

We first show the behavior of $\delta\chi^2$ with $n_{\rm T}$ being
fixed to be the fiducial value; the result is shown on the $T_{\rm R}$
vs.\ $\bar{\Omega}_{\rm IGW}$ plane (top panels).  These figures
indicate that, if we impose $|n_{\rm T}|\ll 1$, $T_{\rm R}$ is always
bounded from below irrespective of the fiducial value of the reheating
temperature (as far as the signal is detected).  We also show the
contours of $\delta\chi^2=5.99$ with $n_{\rm T}$ being marginalized
(middle figure); no prior for $n_{\rm T}$ is imposed in the
calculation for these figures.  We can see that the accuracy of the
determination of fundamental parameters becomes drastically worse if
we marginalize $n_{\rm T}$.  In particular, for the case of
$\hat{T}_{\rm R}=10^7\ {\rm GeV}$, the allowed parameter space extends
to the region of low reheating temperature (and high value of
$\bar{\Omega}_{\rm IGW}$) for some choices of noise parameters.  This
behavior can be understood as follows.  If the postulated value of
$T_{\rm R}$ is lower than the fiducial one, the postulated IGW
spectrum decreases more rapidly than the fiducial one with the
increase of $f$.  Such a discrepancy can be compensated by adopting a
highly blue-tilted IGW spectrum, i.e., large and positive value of
$n_{\rm T}$.  (See also the behavior of $\delta\chi^2$ on the $T_{\rm R}$
vs.\ $n_{\rm T}$ plane.)  To see this, we also show the contours of
the best-fit value of $n_{\rm T}$ to be zero on the $T_{\rm R}$ vs.\
$\bar{\Omega}_{\rm IGW}$ plane (black-dashed lines in middle panels).
We can see that relatively high value of the tensor spectral index
(i.e., $n_{\rm T}\sim 1$) is needed in some region to make the
fiducial and postulated spectra consistent.  This fact implies that,
if a prior for $n_{\rm T}$ is imposed, we may obtain a lower bound on
the reheating temperature even if $n_{\rm T}$ is marginalized.  In
simple slow-roll inflation models, the expansion rate during inflation
decreases with time, which results in a negative value of $n_{\rm T}$.
Concentrating on the parameter region with $n_{\rm T}<0$, for example,
the cases with and without the marginalization of $n_{\rm T}$ give
similar lower bounds on the reheating temperature. We also note here
that, if the reheating temperature is too high, the reheating
temperature is bounded only from below.  (See figures for
$\hat{T}_{\rm R}=10^9\ {\rm GeV}$.)  This is because, with high enough
reheating temperature, $f_{\rm R}$ becomes much larger than $\sim 1\
{\rm Hz}$ so that the shape of the IGW spectrum for the frequency
relevant for the GW detectors becomes almost flat.

The contours of constant $\delta\chi^2$ on the $T_{\rm R}$ vs.\ $n_{\rm
  T}$ plane, with the marginalization of $\log_{10}\bar{\Omega}_{\rm IGW}$, are
also shown (bottom panels); no prior for $\log_{10}\bar{\Omega}_{\rm IGW}$ is
imposed in the calculation for these figures.  Assuming that the
expansion rate during inflation decreases with time, precise
determination of the tensor-to-scalar ratio at the CMB scale imposes
upper bound on $\bar{\Omega}_{\rm IGW}$. In the figures, we also show
the contour on which the best-fit value of $\bar{\Omega}_{\rm IGW}$
becomes equal to $5\times 10^{-16}$, which is the value given by the
tensor-to-scalar ratio at the CMB scale of $0.15$ (which is the
prediction of the chaotic inflation model).

\begin{figure}
  \begin{center}

  \begin{minipage}{0.43\columnwidth}
    \begin{center}
      \includegraphics[clip, width=1.0\columnwidth]{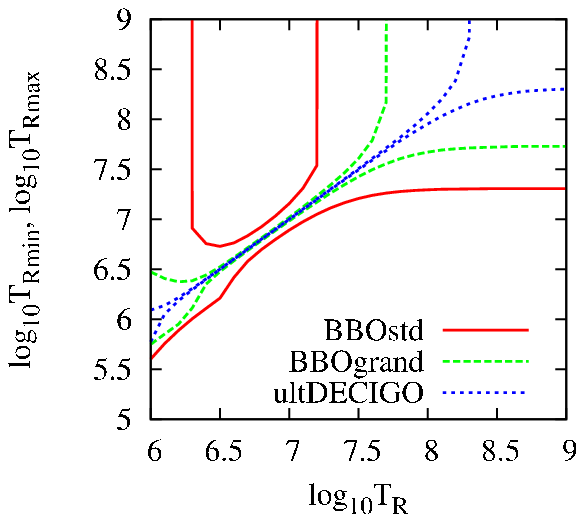}
    \end{center}
  \end{minipage} 
   \begin{minipage}{0.43\columnwidth}
    \begin{center}
      \includegraphics[clip, width=1.0\columnwidth]{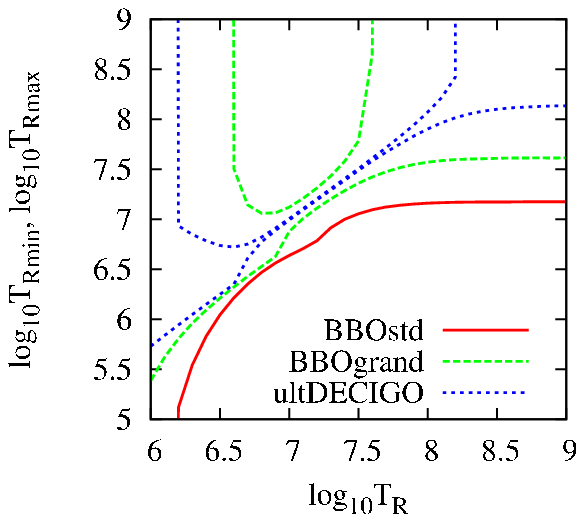}
    \end{center}
  \end{minipage}
  \begin{minipage}{0.43\columnwidth}
    \begin{center}
      \includegraphics[clip, width=1.0\columnwidth]{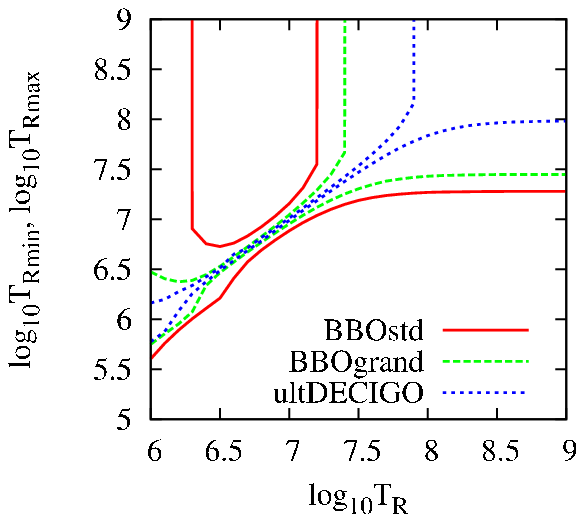}
    \end{center}
  \end{minipage}
    \begin{minipage}{0.43\columnwidth}
    \begin{center}
      \includegraphics[clip, width=1.0\columnwidth]{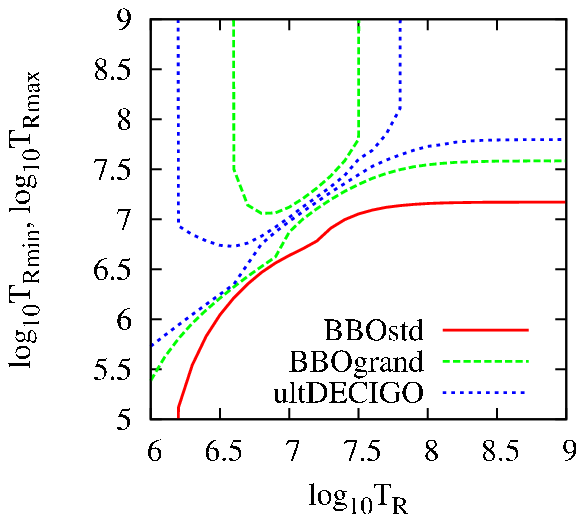}
    \end{center}
  \end{minipage}
  \begin{minipage}{0.43\columnwidth}
    \begin{center}
      \includegraphics[clip, width=1.0\columnwidth]{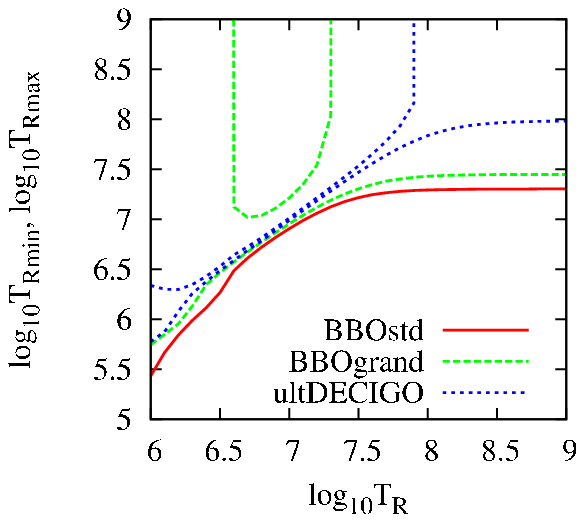}
    \end{center}
  \end{minipage}
  \begin{minipage}{0.43\columnwidth}
    \begin{center}
      \includegraphics[clip, width=1.0\columnwidth]{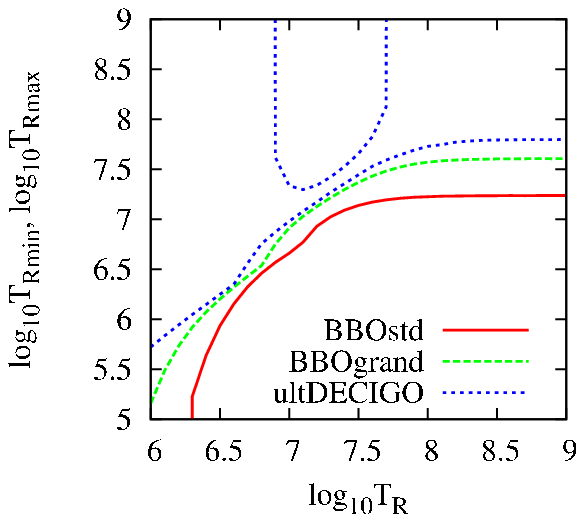}
    \end{center}
  \end{minipage}
  \end{center}
  \caption{\small Upper and lower bounds on $T_{\rm R}$ obtained from
    BBO-std, BBO-grand and ult-DECIGO for $f_{\rm min}=0.1$ Hz (left)
    and $0.3$ Hz (right). In each figure, $\ln \bar{\Omega}_{\rm IGW}$
    is marginalized with a flat prior for $-\infty < \ln
    \bar{\Omega}_{\rm IGW} < -15.3$.  Also, $n_{\rm T}$ is fixed to
    the fiducial value (top), marginalized for $-0.1<n_{\rm T} < 0$
    (middle) and marginalized for $n_{\rm T} < 0$ (bottom).}
  \label{fig:highTR_TRconstraint}
\end{figure}

We can see that the accuracy of the determination of the tensor
spectral index is poor in particular when the reheating temperature is
relatively low (see the figures with $\hat{T}_{\rm R}=10^7\ {\rm GeV}$).
This is because of the suppression of the IGW spectrum for $f\gtrsim
f_{\rm R}$; such a behavior may be mimicked by tilting the spectrum.
However, if the reheating temperature is high enough, the accuracy of
the determination of $n_{\rm T}$ does not change so much (see the
figures with $\hat{T}_{\rm R}=10^9\ {\rm GeV}$).
\footnote{
In Figs.\ \ref{fig:lowTR_BBOstd_TR=10^7} $-$ \ref{fig:lowTR_ulDECIGO_TR=10^9}, 
we do not show the contours which correspond to particular confidence levels. 
This is because of the following reasons. First, in some values of $\hat{T}_{\rm R}$, 
the likelihood function does not converge to $0$ sufficiently for $T_{\rm R} \rightarrow \infty$. 
The contours of particular confidence levels in such cases are sensitive to  
the prior region of $\ln T_{\rm R}$ which we choose, and thus may not be 
reasonable ones. Second, these contours also depends on whether 
$T_{\rm R}$ or $\ln T_{\rm R}$ is considered as the fundamental parameter. 
For the same reason, we define the upper and lower limit of $T_{\rm R}$ 
by the value of $\delta \chi^2$ in Fig.\ \ref{fig:highTR_TRconstraint}.
}
  
We also compare our results with those of Fisher analysis.  For this
purpose, we calculate the Fisher matrix in the parameter space of $\{ p_i
\}=\{ \log_{10} \bar{\Omega}_{\rm IGW}, n_{\rm T}, \ln T_{\rm R} \}$.  
95\ \% C.L. bound with the Fisher analysis is also shown in Figs.\
\ref{fig:lowTR_BBOstd_TR=10^7} $-$ \ref{fig:lowTR_ulDECIGO_TR=10^9} as
pink-shaded regions.  When the fundamental parameters are well
constrained, two analyses give more or less similar bounds.  When the
error in the measurement become sizable, on the contrary, this is not
the case.  In particular, if the fiducial value of the reheating
temperature is so high that $T_{\rm R}$ can be bounded only from
below, constraints from two analyses show significant difference.
This is mainly because the Gaussian approximation discussed in Section
\ref{sec:stat} breaks down when the postulated value of the reheating
temperature becomes much smaller or larger than the fiducial value
(see Figs.\ \ref{fig:lowTR_BBOstd_TR=10^9} $-$
\ref{fig:lowTR_ulDECIGO_TR=10^9}).  Thus, for the precise
determination of the bounds on the reheating temperature, analysis
based on the full likelihood function is suggested.

Finally, we also show the expected accuracy of the determination of
$T_{\rm R}$.  In Fig.\ \ref{fig:highTR_TRconstraint}, we show the
expected upper and lower limits of $T_{\rm R}$ as functions of the
fiducial value of $T_{\rm R}$ for $f_{\rm min}=0.1$ Hz (left) and
$0.3$ Hz (right).  In these figures, the observation time is assumed
to be $T_{\rm obs}=10$ yr.  Also, the upper and lower limits are
defined as the postulated $T_{\rm R}$ which gives $\delta \chi^2 = 4$
after the marginalization of $\ln\bar{\Omega}_{\rm IGW}$ and $n_{\rm T}$.
Here, in order to set the prior for $\ln\Omega_{\rm IGW}(f)$, we adopt
a mild assumption that $\Omega_{\rm IGW}(f)$ decreases with frequency,
which is the case in slow-roll inflation models.  Requiring that
$\Omega_{\rm IGW}(f)$ at the scale relevant for the GW detector does
not exceed that at the CMB scale, we assume a flat prior for $\ln
\Omega_{\rm IGW}(f)$ in $[-\infty,-15.3]$. (The upper bound
corresponds to $\bar{\Omega}_{\rm IGW}=5\times 10^{-16}$.) In
addition, $n_{\rm T}$ is fixed to be the fiducial value in the top
panels while we adopt a flat prior in $[-0.1,0]$ (middle)
or $[-\infty,0]$ (bottom) in other panels. 

From the top-left panels of Fig.\ \ref{fig:highTR_TRconstraint}, one
can see that $T_{\rm R}$ is bounded from both below and above with the
sensitivity of BBO-std for $10^{6.5}$ GeV $\lesssim T_{\rm R} \lesssim
10^{7}$ GeV, if $f_{\rm min}=0.1$ Hz and $n_{\rm T}$ is fixed to the
fiducial value.  With better noise levels like BBO-grand and
ult-DECIGO, such a region for $T_{\rm R}$ becomes broader.  For higher
$\hat{T}_{\rm R}$ only a lower bound is obtained, and this is
consistent with the left panels of Figs.\
\ref{fig:lowTR_BBOstd_TR=10^9} $-$ \ref{fig:lowTR_ulDECIGO_TR=10^9}.
If $|n_{\rm T}|$ is required to be much smaller than $1$, the
qualitative behaviors of the bounds are more or less the same  (see
the case with the marginalization of $n_{\rm T}$ for $[-0.1,0]$).  If
the marginalization is for $[-\infty,0]$, on the contrary, the upper
bound becomes significantly worse.  In particular, the upper bound
cannot be obtained with the noise level of BBO-std.  We can also see
that the bounds are strongly dependent on the minimum frequency
$f_{\rm min}$.  One can see that the bounds with $f_{\rm min} = 0.3$
Hz are much worse than those with $f_{\rm min} = 0.1$ Hz.\footnote
{ In some panels of Fig.\ \ref{fig:highTR_TRconstraint}, one finds
  that analyses with $f_{\rm min}=0.3$ Hz give stronger constraints
  than those with $f_{\rm min}=0.1$ Hz.  This is an artifact of the
  prior region we choose.  For the fiducial value of $T_{\rm R} \simeq
  10^{7.5}$ GeV in the middle figures, for example, the upper bound is
  stronger for $f_{\rm min} = 0.3$ Hz for BBO-grand and ult-DECIGO.
  If one tries to fit the fiducial GW spectrum with $T_{\rm
    R}=10^{7.5}$ GeV using a postulated GW spectrum with postulated
  value of $T_{\rm R}$ higher than the fiducial $T_{\rm R}$, the
  best-fit value of $n_{\rm T}$ is smaller (the absolute value is
  larger) for $f_{\rm min} = 0.3$ Hz, and therefore such a postulated
  value of $T_{\rm R}$ tends to be rejected for $f_{\rm min}=0.3$ Hz.
  If $n_{\rm T}$ is fixed to the fiducial value (top figures), there
  is no such an artifact.  }

\section{Testing Chaotic Inflation}
\label{sec:test}
\setcounter{equation}{0}

With the determination of $\bar{\Omega}_{\rm IGW}$, $n_{\rm T}$ and
$\alpha_{\rm T}$ (as well as $T_{\rm R}$), information about the
properties of inflaton may be obtained.  Thus, we briefly discuss the
implication of the determination of the IGW spectrum for the test of
inflation models.  If we take the chaotic inflation model we have
introduced in Section \ref{sec:spectrum} as the model of inflation,
the properties of the inflaton can be parameterized by two parameters,
the inflaton mass $m_\phi$ and the decay rate $\Gamma_\phi$ (or
equivalently, the reheating temperature).  If the reheating
temperature is in the relevant range, it may be directly determined
from the study of the IGW spectrum, as we have explained in the
previous section.  In such a case, we can determine one of the very
important parameter, $T_{\rm R}$.  Thus, in this section, we consider
the case where $T_{\rm R}$ is very high; even in such a case, we will
see that we may have a chance to acquire the upper and lower bounds on
$T_{\rm R}$.

As we have discussed, the reheating temperature is as high as
$10^{10}\ {\rm GeV}$ if the inflaton decays via a Planck-suppressed
operator.  In such a case, the IGW spectrum at $f\sim 1\ {\rm Hz}$ is
insensitive to the reheating temperature, and hence the procedure to
determine $T_{\rm R}$ discussed in the previous section is difficult.
As shown in Table \ref{table:params}, however, $\bar{\Omega}_{\rm
  IGW}$, $n_{\rm T}$ and $\alpha_{\rm T}$ have slight dependences on
the reheating temperature (if the value of $m_\phi$ is fixed).  This
is because the number of the $e$-folding during the inflation varies
as the reheating temperature changes.  Then, the inflaton amplitude at
the time of the horizon exit of the mode $f_*$, which is denoted as
$\phi_*$, changes.

In the chaotic inflation model with the inflaton potential given in
Eq.\ \eqref{infpot}, the expansion rate and the slow-roll parameters
at the time when the mode $f_*$ exits the horizon are given by
\begin{align}
  H_* = \frac{m_\phi \phi_*}{\sqrt{6}M_{\rm Pl}},~~~
  \epsilon_* = \eta_* = 2 \frac{M_{\rm Pl}^2}{\phi_*^2}.
\end{align}
Thus, combining these relations with Eqs.\ \eqref{BarOmg} and
\eqref{n_T}, the values of $m_\phi$ and $\phi_*$ may be obtained with
the measurement of $\bar{\Omega}_{\rm IGW}$ and $n_{\rm T}$.  Because
the value of $\phi_*$ (and hence $n_{\rm T}$) is insensitive to
$m_\phi$, information about $n_{\rm T}$ can be translated to that
about the reheating temperature.  As shown in Table
\ref{table:params}, the predicted value of $n_{\rm T}$ varies from
$-0.0710$ to $-0.0581$ for $T_{\rm R}=10^{7}-10^{12}\ {\rm GeV}$.
Thus, the error in the measurement of $n_{\rm T}$ should be at the
level of $O(10^{-3})$ in order to acquire sensible information about
the reheating temperature.  The noise level slightly better than that
of BBO-grand is required in order to perform such an analysis.  If the
noise level of ult-DECIGO is available, on the contrary, $n_{\rm T}$
can be precisely determined.  In such a case, assuming the chaotic
inflation, $T_{\rm R}$ is well determined even if $f_{\rm R}$ is out
of the sensitivity range of the GW detectors.  For example, assuming
that $n_{\rm T}$ is measured as $n_{\rm T}=-0.0639\pm 0.00061$ and
$n_{\rm T}=-0.0639\pm 0.00093$, which are the expected accuracy in the
two- and three-parameter analysis with $f_{\rm min}=0.1\ {\rm Hz}$,
respectively (see Table \ref{table:1sigma}), the reheating temperature
is estimated to be $(0.6 - 1.5)\times 10^{10}\ {\rm GeV}$ and $(0.5 -
1.9)\times 10^{10}\ {\rm GeV}$, respectively.

\section{Conclusions and Discussion}
\label{sec:conclusions}
\setcounter{equation}{0}

We have discussed the prospects of the measurement of IGW spectrum
using future space-based GW detectors, like BBO and DECIGO.  We have
performed a detailed analysis for the determination of IGW parameters,
i.e., the amplitude $\bar{\Omega}_{\rm IGW}$, the tensor spectral
index $n_{\rm T}$, and its running $\alpha_{\rm T}$.  We have adopted
the chaotic inflation model with parabolic inflaton potential as a
fiducial model, and calculated the IGW amplitude for the frequency
relevant for the IGW detectors.  Then, using such an amplitude as well
as tensor spectral index and running as the fiducial values, we
performed a statistical analysis to estimate the expected accuracy of
the measurements of these parameters. Here, we considered two cases.
One is the case with high enough reheating temperature $T_{\rm R}$,
for which the IGW spectrum becomes insensitive to $T_{\rm R}$.  The
other is the case where $T_{\rm R}$ is so low that the IGW detectors
may directly observe the signal of the reheating; in such a case, the
which is the IGW spectrum is significantly suppressed in high
frequency region.

In the case with high enough reheating temperature, we have shown
expected accuracies of the measurements of $\bar{\Omega}_{\rm IGW}$,
$n_{\rm T}$, and $\alpha_{\rm T}$, adopting several noise parameters
(which we call BBO-std, BBO-grand, and ult-DECIGO).  Adopting the
chaotic inflation model which predicts $r\sim 0.15$ at the CMB scale,
we have seen that non-zero value of the tensor spectral index $n_{\rm
  T}$ may be confirmed with the noise level of BBO-grand with a few
years of operation.  For the detection of the running of the tensor
mode, on the contrary, significant improvement of the noise level is
necessary if $\alpha_{\rm T}\sim O(10^{-3})$.

If the reheating temperature is relatively low, on the contrary, the
future space-based GW detectors may put lower and upper bounds on
$T_{\rm R}$.  We have seen that, if the fiducial value of the
reheating temperature, $\hat{T}_{\rm R}$, is $\sim 10^{6.5}-10^{7.5}\
{\rm GeV}$, the reheating temperature can be well constrained.  We
have estimated the expected bounds on the reheating temperature.  In
particular, with the two-parameter analysis taking $\bar{\Omega}_{\rm
  IGW}$ and $T_{\rm R}$ as free parameters, we have seen that the
reheating temperature can be determined with the error of $\sim 30\
\%$ with BBO-std and $\sim 5\ \%$ with BBO-grand if $\hat{T}_{\rm
  R}\sim 10^7$ GeV, assuming 10 years of operation and the minimum
frequency $f_{\rm min}=0.1$ Hz.  If $\hat{T}_{\rm R}\gtrsim 10^8 {\rm
  GeV}$, on the contrary, the reheating temperature is bounded only
from below.  We have also compared our results with full likelihood
with those with Fisher matrix analysis.  We have seen the results of
two analysis may differ significantly in some cases.

The determination of the tensor-to-scalar ratio provides important
information about the normalization of the cosmic IGW background.
Although it may be premature to conclude that $r\sim O(0.1)$ based
only on the result of BICEP2, our knowledge about the $B$-mode signal
will be significantly improved in the near future because many efforts
to detect the $B$-mode signal in CMB are on-going. Once the existence
of the IGWs is confirmed by the observation of the $B$-mode signal in
the near future, the program to detect and study the IGW spectrum in
future space-based GW detectors is strongly suggested.  Such a program
will provide important and unique information about inflation.

\vspace{1em}

\noindent {\it Note Added}

\noindent
While we are finalizing this manuscript, the paper
\cite{Kuroyanagi:2014qaa} appeared on the arXiv, which may have some
overlap with our analysis.

\vspace{1em}

\noindent {\it Acknowledgements}

\noindent
The authors are grateful to M. Ando for valuable discussion. 
They appreciate useful comments by J.~'I.~Yokoyama. 
They also thank A. Taruya for correspondence concerning their works.  
One of the authors (T.M.) is grateful to the Mainz Institute for
Theoretical Physics (MITP) for its hospitality and its partial support
during the completion of this work.
This work is supported by JSPS KAKENHI Grant No.~26400239 (T.M.),
No.~60322997 (T.M.), No.~23740195 (T.T.).  R.J. is supported by the
JSPS fellowship (No.~25-8360) and also by Program for Leading Graduate
Schools, MEXT, Japan.



\end{document}